\documentclass[preprint2]{aastex}
\usepackage{color}
\usepackage{rotating}
\shorttitle{Convection and mode suppression in red giants in NGC\,6811}
\shortauthors{T. Arentoft et al.}

\begin{document}
\title{Convective-core overshoot and suppression of oscillations: Constraints from red giants in NGC\,6811}
\author{T. Arentoft\altaffilmark{1}, K. Brogaard, J. Jessen-Hansen, V. Silva Aguirre, H. Kjeldsen \\
and J.\,R. Mosumgaard} 
\affil{Stellar Astrophysics Centre, Department of Physics and Astronomy, Aarhus University, \\ Ny Munkegade 120, DK-8000 Aarhus C, Denmark}
\author{E.\,L. Sandquist} 
\affil{San Diego State University, Department of Astronomy, San Diego, CA 92182, USA}
\altaffiltext{1}{toar@phys.au.dk}

\begin{abstract}
Using data from the NASA spacecraft {\it Kepler}, we study solar-like oscillations in red-giant stars in the open cluster NGC\,6811.
We determine oscillation frequencies, frequency separations, period spacings of mixed modes and mode visibilities for eight cluster giants. 
The oscillation parameters show that these stars are helium-core-burning 
red giants. The eight stars form two groups with very different oscillation power spectra; the four stars with lowest $\Delta\nu$-values display rich sets of mixed $l=1$ modes, 
while this is not the case for the four stars with higher $\Delta\nu$. For the four stars with lowest $\Delta\nu$, we determine the asymptotic period spacing of the mixed modes, $\Delta$P, 
which together with the masses we derive for all eight stars suggest that they belong to the so-called secondary clump.
Based on the global oscillation parameters, we present initial theoretical stellar modeling which indicate that we can constrain convective-core overshoot on the main sequence and in the 
helium-burning phase for these $\sim$2\,M$_{\odot}$ stars. Finally, our results indicate less mode suppression than predicted by recent theories for magnetic suppression of certain oscillation modes 
in red giants. 
\end{abstract}


\section{Introduction}
The NASA spacecraft {\it Kepler} \citep{2010Sci...327..977B} has provided photometric time-series of unprecedented quality for solar-type and red-giant stars, and asteroseismic analysis 
of these data have led to remarkable results as reviewed by \citet{2013ARA&A..51..353C}. The asteroseismic analysis allows for a determination of the evolutionary stage of evolved stars, 
as period spacings of mixed oscillation modes (especially $l=1$) separate hydrogen-shell-burning and helium-core-burning red giants \citep{2011Nature..471..608,2011AA...532..A86,2013ApJ...765L..41S}.
Mixed modes are non-radial modes which behave both like $p$-modes, probing the outer layers of the star, and like $g$-modes, probing the deep interior and hence the evolutionary stage of the star.  

\citet{2013ApJ...765L..41S} showed that $\Delta$P-$\Delta \nu$ (period spacing, large frequency separation) diagrams using both the median period 
spacing, derived from the observed oscillation spectra, and the asymptotic period spacing from theoretical models, separate red giants of different mass and evolutionary stage. 
The evolution of the asymptotic period spacing as a function of mass and age depends on the amount of convective-core overshoot,
which means that period spacings of evolved stars offer the opportunity of constraining theoretical stellar models, including the amount of convective-core 
overshoot on the main sequence for 1.2--4.0\,M$_{\odot}$ stars \citep{2013ApJ..766..118,2015MNRAS..452..123}.  

It has also been found that non-radial modes are suppressed in a subset of the red giants observed with {\it Kepler} \citep{2012A&A...537A..30M}. This has recently been suggested to be associated
with strong internal magnetic fields \citep{2015Sci...350..423F, 2016Natur.529..364S, 2016PASA...33...11S}, which means that asteroseismology of red giants might also offer the opportunity of 
investigating magnetic fields within giant stars.
 
Evolved stars in open clusters observed with {\it Kepler} have been studied using asteroseismology by several groups, e.g., \citet{2011ApJ...739...13S,2012ApJ..757..190,2012MNRAS..419..2077}. 
NGC\,6811 is a cluster of intermediate age ($\sim$\,1\,Gyr) which contains pulsating stars, binaries and red giants, as discussed in detail by \citet{2014MNRAS.445.2446M}.
The presence of both oscillating red giants and detached eclipsing binaries in the same cluster, sharing common parameters such as age, distance and metallicity, offers unique possibilities 
for testing theoretical stellar models.

As described in \citet{2016AN..XX..XX}, our long-term aim is to combine asteroseismology of giant stars with information derived from detached eclipsing binaries, in order to test 
asteroseismic scaling relations and challenge stellar models. The asteroseismic scaling relations relate the global properties of solar-like oscillations (frequency of maximum power $\nu_{\rm max}$ 
and large frequency separation $\Delta \nu$) to stellar mass, radius and effective temperature (see Sect.~6). They are used for deriving stellar properties of (faint) stars for which only this 
global asteroseismic information is available and are therefore important to test.

In this paper we present an analysis of {\it Kepler} light-curves of eight oscillating red giants in NGC\,6811 to derive the global oscillation parameters to be used for testing the asteroseismic
scaling relations, as presented in Sections\,2--5. From the asteroseismic parameters we derive masses and evolutionary stages 
for these stars, we present comparisons of asymptotic period spacings $\Delta$P with initial stellar modeling, and we use those period spacings to show that convective-core overshoot
both on the main sequence and in the helium-core-burning phase can be constrained for these stars using the asteroseismic parameters (Sections 6 and 7). 
Another main result is that we find that the power spectra of the eight stars show very different oscillation properties despite having very similar stellar properties.
We discuss in Sect.\,8 if this could be explained by the magnetic effects mentioned above.

\section{The data and the target stars}
We used {\it Kepler} long-cadence data \citep[$\sim$30 min. sampling,][]{2010ApJ...713L..79K} to study eight red giants in the open cluster NGC\,6811. Five stars (KIC9716522, 9655101, 9655167,
9534041, 9716090) are listed as red giants in \citet{2011ApJ...739...13S}, KIC9532903, 9776739 had their first asteroseismic study by \citet{2011A&A...530A.100H} and were identified as cluster members by 
\citet{2014MNRAS.445.2446M}. KIC9409513 was found in our own investigations of the cluster, and is also included in \citet{Sand2016..831..11S} who furthermore show that it is unlikely to find more 
cluster-member giants. The giants have $V\sim11$ allowing spectroscopic studies \citep{2014MNRAS.445.2446M}.

Each time series includes more than 60,000 data points and spans $\sim$1400 days (BJD 2454953 - 2456391, {\it Kepler} quarters Q1 -- Q17) from data-release 22. The data
were extracted from {\it Kepler} pixel-data using reduction software kindly provided to us by S. Bloemen. As the standard {\it Kepler} light-curves are not optimized for photometry in crowded 
fields, as in an open cluster, we manually defined large pixel apertures for each star in order to optimize the light curves. 
The extracted light-curves were filtered with the KASOC filter \citep{2014MNRAS.445.2698H} to prepare the data for asteroseismic analysis. The filter 
operates with two timescales, one long and one short, to remove trends and signals that do not originate from the stellar oscillations. To test the robustness of the filtering, we have 
worked with two versions of the filtered data, one with long and short timescales of 30\,d and 3\,d, and one with 10\,d and 1\,d, see \citet{2014MNRAS.445.2698H} for details. Quantitative comparisons 
show that the power spectra of the two filtered versions differ noticeably only at very low frequencies (below $2\mu$Hz). We do not use this very low-frequency part of the power spectra in 
our analysis and we find consistent results using the two versions of the time series, as the results of our asteroseismic analysis differ only within the 1\,$\sigma$ uncertainties. 
We adopt the mean values from the two versions as our final values.

\begin{figure*}
\center \includegraphics[width=145mm]{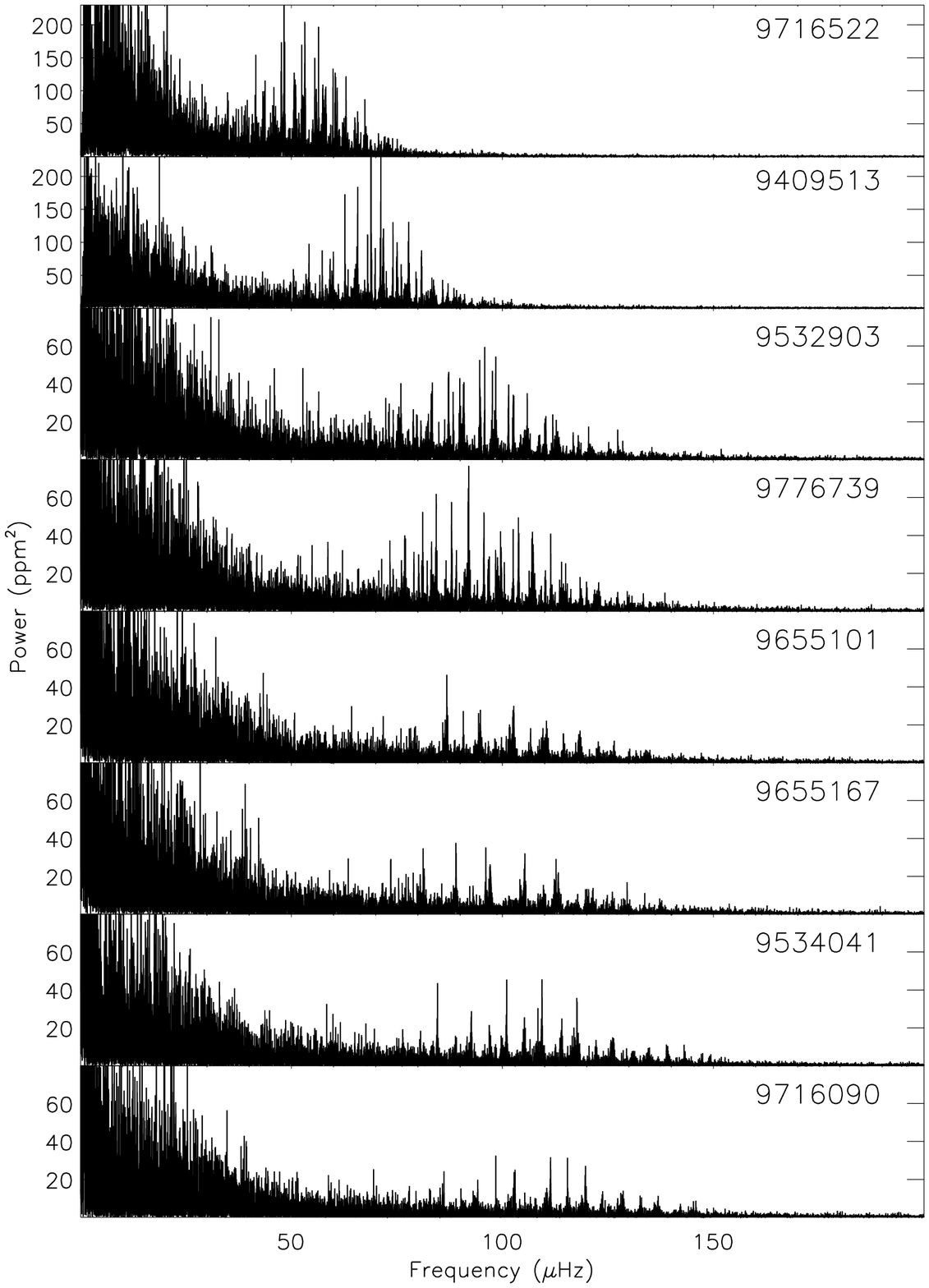}
 \caption{Power spectra of the eight stars included in this study, arranged in the same order as in Table~\ref{tab.1}. Notice that the two top panels have y-scales different from the lower 6 panels. Remarkable
differences in amplitude and richness of modes are seen between the top four and the bottom four power spectra.}
\label{fig.power}
\end{figure*}

The list of analyzed stars is found in Table~\ref{tab.1} and the power spectra of their light curves are presented in Fig.~\ref{fig.power}. The stars are arranged in order of increasing large frequency
spacing $\Delta\nu$ and frequency of maximum power, $\nu_{\rm max}$, in both Table~\ref{tab.1} and Fig.~\ref{fig.power}. We will describe how we determine these 
asteroseismic parameters below, however already at this point we notice the remarkable feature in the power spectra of the eight stars shown in Fig.~\ref{fig.power};
the top four stars all have higher amplitudes and richer oscillation spectra as compared to the four bottom stars. We return to this below. 

\section{The stellar background}
The stellar backgrounds originating from convection (granulation, activity) were fitted in the power spectra using a two-component model \citep{handberg16},
including the attenuation factor $\eta(\nu)$ which is needed with {\it Kepler} long-cadence data \citep{2014A&A...570A..41K}, of the following form:
\begin{small}
\begin{equation}
N(\nu) = \eta(\nu) \cdot \sum_{k=1}^{2} \frac{4\sigma_{k}^{2}\tau_{k}}{(1 + (2\pi\nu\tau_{k})^{2})^{2}} + K.
\end{equation}
\end{small}
Here, $\eta(\nu) = sinc^{2}(\Delta T_{int} \cdot \nu)$ and $\Delta T_{int} = 1765.5$\,s for the {\it Kepler} long-cadence data. $K$ is the white-noise level, $\nu$ is the frequency, 
and $\tau_k$ and $\sigma_k$ are the timescales and amplitudes of each component. 

\begin{figure*}
\center \includegraphics[width=150mm]{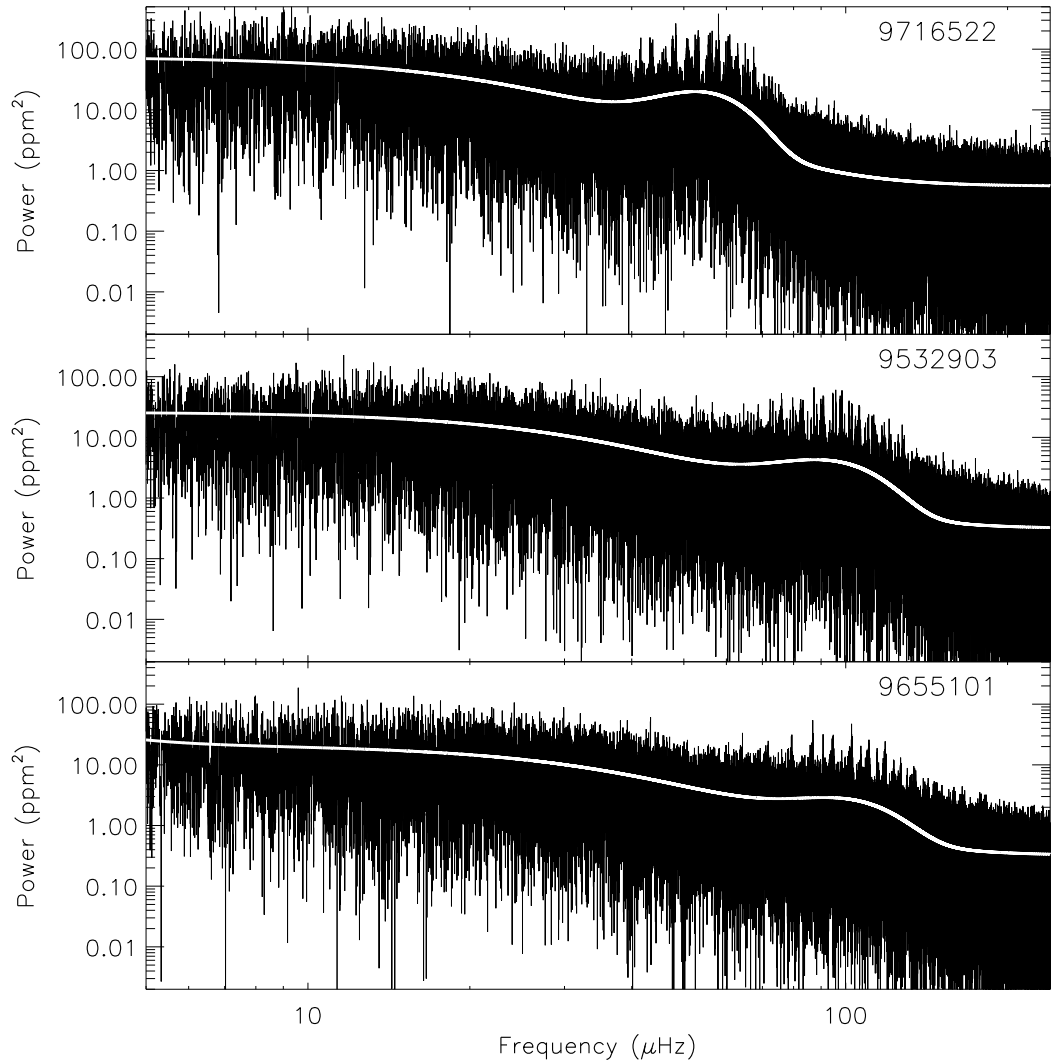}
 \caption{Examples of power spectra with the fit to the stellar activity background and the oscillation envelope superimposed. The three stars shown are KIC9716522, KIC9532903 and KIC9655101, and
the input data used for the plots are the 10/1\,d filtered data (see text). Plots using the 30/3\,d filtered data would look very similar.
The frequencies of maximum power, $\nu_{\rm max}$, are 53.7$\pm$1.0\,$\mu$Hz, 92.0$\pm$1.5\,$\mu$Hz and 98.7$\pm$1.0\,$\mu$Hz, respectively (see Table~\ref{tab.1}).}
\label{fig.loglog}
\end{figure*}

We include a Gaussian envelope to account for the oscillation signal and to determine the frequency of maximum power ($\nu_{\rm max}$), such that the power spectrum is modelled 
using this expression \citep{handberg16}:
\begin{small}
\begin{equation}
P(\nu) = N(\nu) + \eta(\nu) \cdot a_{\rm env} \cdot \rm{exp} \left( \frac{-(\nu -\nu_{\rm max})^{2}}{2\sigma_{\rm env}^{2}} \right).
\end{equation}
\end{small}
Here, $a_{\rm env}$ and $\sigma_{\rm env}$ are the height and width of the envelope.
Three examples of power spectra with their background fits and Gaussian envelopes superimposed are shown in Fig.~\ref{fig.loglog}.

It is known that fitting the stellar background is a difficult task, see e.g. \citet{handberg16} for a discussion. We have tried different versions of the background fit, but 
found the model above to give the best results. This is in agreement with \citet{handberg16} who empirically, on the basis of extensive statistical tests on a number of red giants in the open cluster
NGC\,6819 observed with {\it Kepler}, found this model to be optimal for stars with $\nu_{\rm max}$ in the range of our stars. 
We found that the value of $\nu_{\rm max}$ is affected by the choice of background model and may differ by several $\mu$Hz from one model to the other. The difference is within the uncertainties for most
stars, but for the two stars in Table~\ref{tab.1} with $\nu_{\rm max}$ above 100\,$\mu$Hz, differences in $\nu_{\rm max}$ between background models were up to 5\,$\mu$Hz. We trust the model 
above (Eq.~1) for three reasons; 1) it is found
by \citet{handberg16} to be optimal for this type of stars, 2) the fit converges better (faster) than for the other models and 3) the results show the expected correlation between $\Delta\nu$ and 
$\nu_{\rm max}$, see for example \citet{2009MNRAS.400L..80S}, which is not the case when we use other models.

\section{The asteroseismic analysis}

The frequency of maximum power $\nu_{\rm max}$ was determined for the eight stars using the fitting procedure described in the previous section. 
To estimate the uncertainty on $\nu_{\rm max}$ we split each time series (two time series for each star, filtered using two different timescales as described above) in two halves, fitted each half 
separately using Eq.~2, and used the largest differences between the values of $\nu_{\rm max}$ 
derived from the fit to the full series and from the fits to the half series for our uncertainty estimates. The estimates were taken as the largest difference divided by $\sqrt2$, however in some 
cases the derived uncertainty is below 1.0\,$\mu$Hz which seems optimistic when comparing to \citet{Sand2016..831..11S} who also determined $\Delta\nu$ and $\nu_{\rm max}$ for these stars, using an 
automated pipeline.
We therefore applied a lower limit on the uncertainty on $\nu_{\rm max}$ of 1.0\,$\mu$Hz. For a couple of the stars we did try to split the time series in four instead of two 
and found consistent results whether we used the same method as above or the standard deviation of the mean value. 
The results are shown in Table~\ref{tab.1}. We note that our $\nu_{\rm max}$ values generally compare well with those of \citet{Sand2016..831..11S} except for KIC9534041, where
the errorbars overlap at the 2$\sigma$ level only. For KIC9655167 \citet{Sand2016..831..11S} quotes a higher $\nu_{\rm max}$ value than ours, but also a larger error of nearly 6.0\,$\mu$Hz. We finally note
that our uncertainty values are comparable to those of \citet{handberg16} for similar stars in NGC\,6819.

Solar-like oscillations in red giants follow approximately the asymptotic relation \citep{vandakurov1967,tassoul1980,gough1986} 

\begin{small}
\begin{equation}
\nu_{n,l}\approx\Delta\nu(n+\frac{1}{2}l+\epsilon)-l(l+1)D_0,
\end{equation}
\end{small}

where $\nu$ is the frequency, $n$ is the radial order of the oscillations and $l$ is the angular degree, $\Delta\nu$ is the large frequency separation between modes with the same $l$-value, 
$\epsilon$ is a dimensionless 
parameter which is sensitive to the surface layers, and $D_0 = \frac{1}{6}\delta_{02}$, where $\delta_{02}$ is the small frequency separation between modes of $l=0$ and $l=2$. 

To determine the large frequency separation $\Delta \nu$, we first applied a method described in \citet{2008JPhCS.118a2039C}, which was developed with the aim of 
analyzing solar-like oscillations in {\it Kepler} data. For a range of trial $\Delta \nu$-values, the region of the power spectrum containing the oscillations is for each $\Delta \nu$-value cut in 
sections of $\Delta \nu$/2. These sections are then stacked and the highest peak in the summed spectrum is found. This peak will be at its maximum height when the correct $\Delta \nu$ value is used, as the
$l=0,1$ modes in this case will add up and create a single, very strong peak in the summed spectrum. In principle, the mixed $l=1$ modes could bias the results. However, our later refinement using
only $l=0$ frequencies showed that this was not the case.

We then found the frequencies for the individual oscillation modes following a method largely based on \citet{mosumgaard14}, and detected in this way between 28 and 51 frequencies in the eight 
analyzed stars. Solar-like oscillations are stochastic which means that each individual oscillation mode is represented by a number of peaks in the 
power spectrum, together forming a Lorentzian shape. Following \citet{mosumgaard14}, the power spectra were smoothed using $IDL$'s smooth function, taking care that the smoothing 
was as mild as possible to ensure that the mean positions of the oscillation modes remained unaffected. In the smoothed power spectra, we detected the oscillation frequencies using an automated procedure 
where each frequency value in the end, as an extension of the method presented in \citet{mosumgaard14},
was determined from a Gaussian fit to the top part of the corresponding peak in the power spectrum. Each oscillation frequency was then manually inspected, using for each star the power spectra from 
both filtered versions of the time series. 

\begin{figure*}
\center \includegraphics[width=150mm]{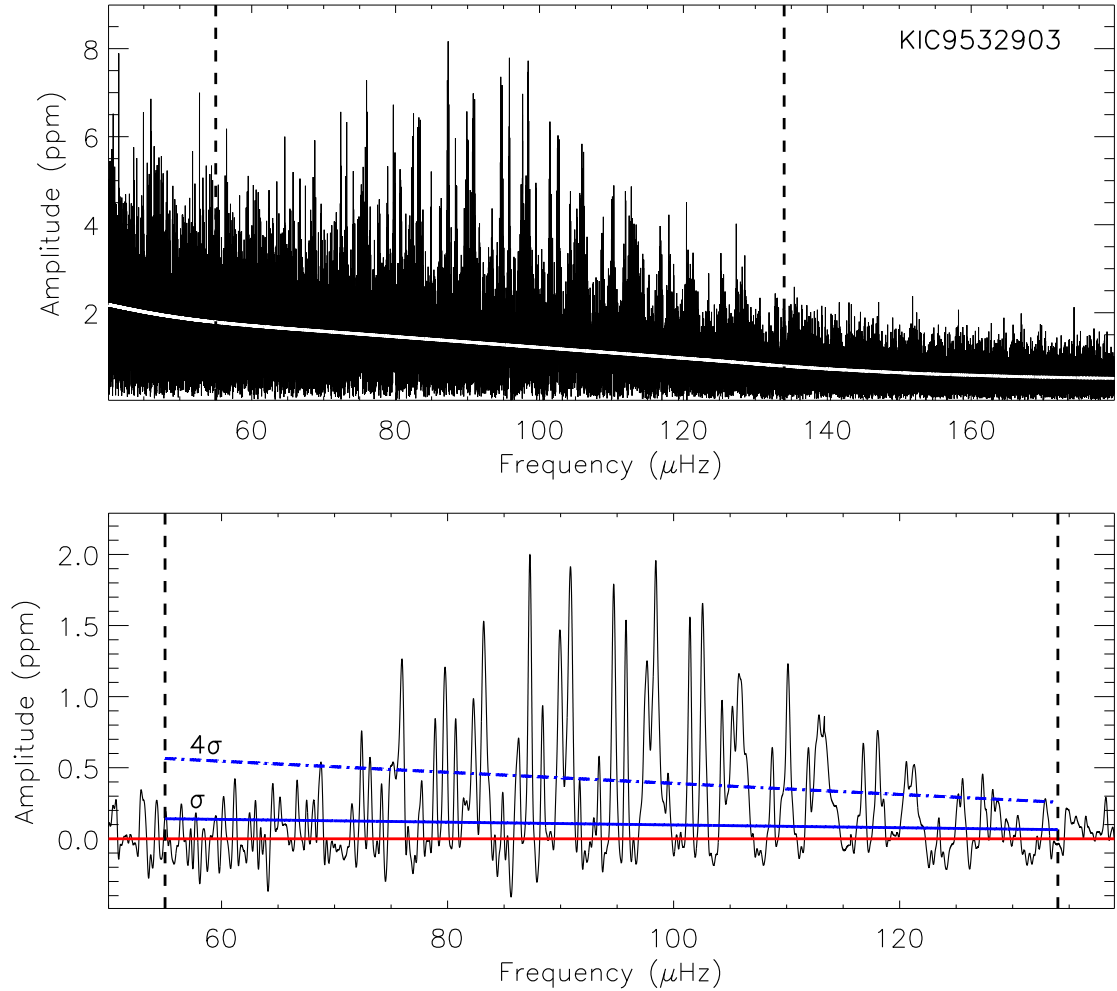}
 \caption{
Frequency detection and signal-to-noise estimation for the oscillations in KIC9532903. The upper panel shows the raw amplitude spectrum with the region of the oscillations bracketed by the vertical 
dashed lines, and the estimated background level shown as the white solid line. The lower panel shows a zoom of the oscillations after subtracting the background signal and smoothing the signal, 
with the solid red line 
as a zero line to guide the eye. The noise levels are estimated to be 0.141 ppm on the low-frequency side of the oscillations and 0.064 ppm on the high-frequency side. 
The solid blue line shows the (linear) frequency dependent noise level in the region of the oscillations, while the dashed-dotted line shows the same line multiplied by 4, illustrating the detection 
limit for statistically significant oscillation peaks.
}
\label{fig.SN}
\end{figure*}

The signal-to-noise ($SN$) values of the individual modes were estimated from the amplitude spectra as illustrated in Fig.~\ref{fig.SN}. 
For each star, we estimated the background in the amplitude spectrum using median boxcar smoothing
on the low- and high-frequency side of the oscillation signal, and used linear interpolation between the low- and high-frequency parts to estimate the background in the region of 
the spectrum where the oscillations are found, see 
Fig.~\ref{fig.SN}.  We subtracted the background signal and determined the amplitudes of all peaks in the residual spectrum, and used the median peak amplitude at the low- and high-frequency side 
to estimate the typical amplitude of a noise peak in these two frequency regions. As the noise is higher at low frequencies than at high frequencies, we again used linear interpolation to estimate the 
noise at the position of a given peak in the region of the oscillations. The $SN$-value was then calculated as the ratio between the amplitude of the suspected oscillation peak and the noise 
estimate at the given frequency.

We estimated the uncertainties on the frequencies by first dividing each time series in two halves and determined frequency values from these different spectra. 
As we have two versions of the time series for each star, we obtain four estimates for each frequency, and took the {\it rms}-scatter of these four values as our first guess of the uncertainty  
on the given frequency. We then used the fact that we have a number of detected frequencies for each star; we expect that the frequency uncertainty correlates with the reciprocal 
value of the $SN$. We therefore plotted the frequency uncertainties as a function of $(1/SN)$, fitted a line, and used that line to estimate frequency uncertainty at a given $SN$. 
We have compared the uncertainties derived in this way to those of \citet{handberg16},
who use {\sc MCMC}-techniques on similar data for similar stars in the open cluster NGC\,6819, and found the results to be quite comparable.

We note that the search for frequencies in these stars is not a blind search, as the frequency peaks follow Eq.\,3, at least approximately. In the process of searching for the frequencies, 
we took advantage of Eq.\,3, in the sense that the $l=0,2$ modes form a regular pattern, providing additional support for modes with low $SN$. A small number of the retained modes are 
therefore not significant when seen as isolated peaks, having $SN$-ratios in amplitude of $\sim2$ or even a bit lower, however they fit in the expected mode structure and 
are therefore included, although a few of them may be noise peaks. In Table~\ref{tab.2} we list the frequencies (with uncertainties), $SN$-values and mode-identification for one of the stars,
KIC9532903. The frequencies for the remaining seven stars can be found in the online appendices and in the results section of the {\sc KASOC} data base\footnote{http://kasoc.phys.au.dk/results/}. 

\begin{figure*}
\center \includegraphics[width=150mm]{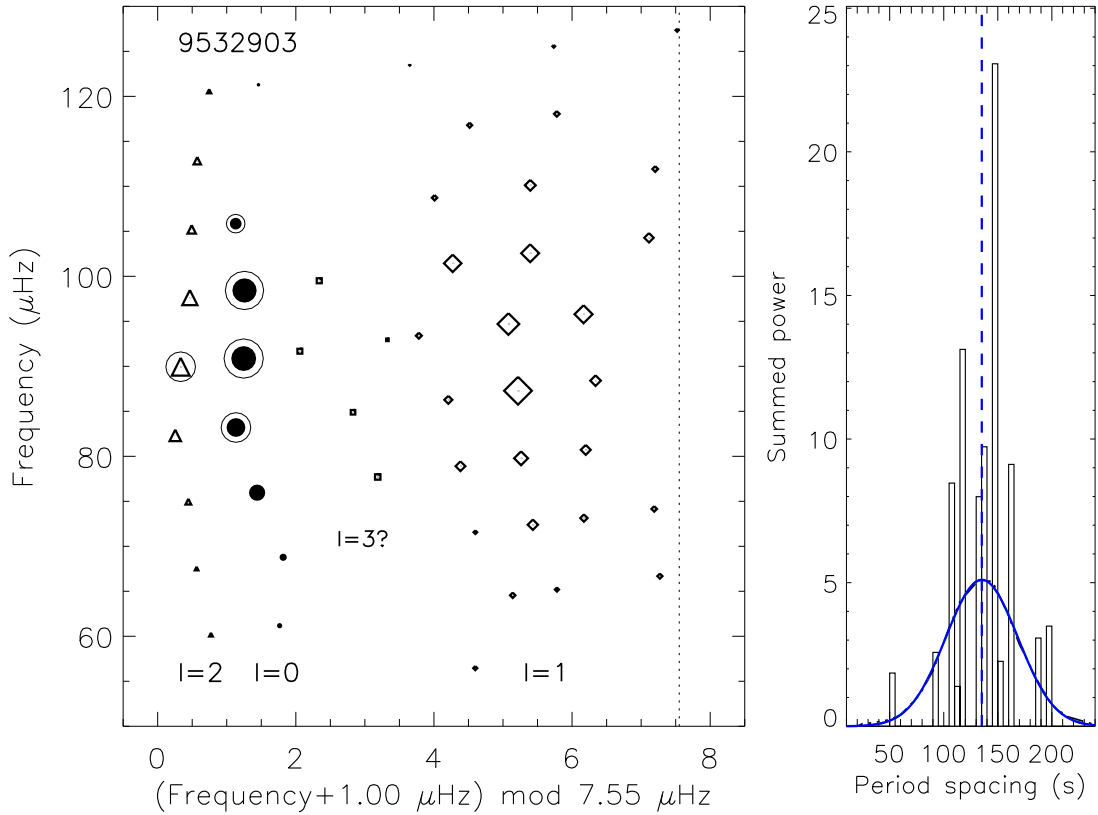}
 \caption{Left: {\'E}chelle diagramme of the detected oscillation frequencies modulo the large frequency separation $\Delta\nu$ on the x-axis, and the oscillation frequencies themselves on the 
y-axis. Ridges of different spherical harmonic degree $l=0-2$ and possibly $l=3$ are identified, and the individual modes are shown as filled circles for $l=0$, diamonds for $l=1$, triangles 
for $l=2$ and squares for possible $l=3$-modes. The ridges have been shifted on the x-axis in the same way as in Fig.~\ref{fig.allechelle}, in order to allign all eight stars in that figure. 
The sizes of the symbols correspond to the amplitudes of the oscillation modes. 
The complicated mode structure for $l=1$ is due to the presence of mixed modes, which are used for determining the period spacing $\Delta$P$_{\rm obs}$. The four encircled $l=0$ modes are used for determining
the large frequency separation $\Delta\nu$, and the encircled $l=2$ mode is used, together with the neighboring $l=0$ mode, for determining the small frequency separation $\delta_{\rm 02}$.
Right: Histogram of period differences for the $l=1$ modes. This plot is described in the text.}
\label{fig.echelle}
\end{figure*}

For each star, we then constructed {\'e}chelle diagrams using the first value of the large frequency separation determined above. 
An example of such an {\'e}chelle diagramme is shown in the left panel of Fig.~\ref{fig.echelle}. In the {\'e}chelle diagramme, the oscillation modes are separated along the x-axis into ridges according to 
the degree 
($l$-value) of the modes, and along the y-axis according to the radial order ($n$-value) of the modes. In this diagramme, modes of $l=0-3$ can therefore be identified. We are particularly interested
in the mixed-mode structure at $l=1$, which we will return to in the following section. Here, we mention that the final value of the large frequency separation is determined from a weighted linear fit to 
the four central $l=0$ modes, using the uncertainties on the individual frequencies as weights and for the final error estimate, see Table~\ref{tab.1}. 

Although we only have measurements for eight stars in NGC\,6811, we can still investigate the relation between $\Delta\nu$ and $\nu_{\rm max}$. Fitting a power law to the data, we find the 
following relation:
\begin{equation}
\Delta\nu = (0.193\pm0.003) \cdot \nu_{\rm max}^{(0.812\pm0.003)}.
\end{equation} 

The coefficients found here are slightly different from those of other studies, e.g. \citet{2009MNRAS.400L..80S}, or \citet{handberg16} who found a constant of $0.248\pm0.009$ and a power
of $0.766\pm0.008$ for the giants in NGC\,6819. However, the difference is in line with expectations since the relationship between $\Delta\nu$ and $\nu_{\rm max}$ is mass-dependent
\citep{2010A&A...517A..22M,2010ApJ...723.1607H}.

The small frequency separation ($\delta_{\rm 02}$) is found from the central $l=0,2$ modes, see the left panel of Fig.~\ref{fig.echelle}. 
Finally the value of $\epsilon$ is determined from the $l=0$ modes using Eq.~3. These values are listed for completeness in Table~\ref{tab.1}, along with surface gravities determined from the asteroseismic 
parameters, but are not used for the analysis in the present paper.    

\begin{figure*}
\center \includegraphics[width=118mm]{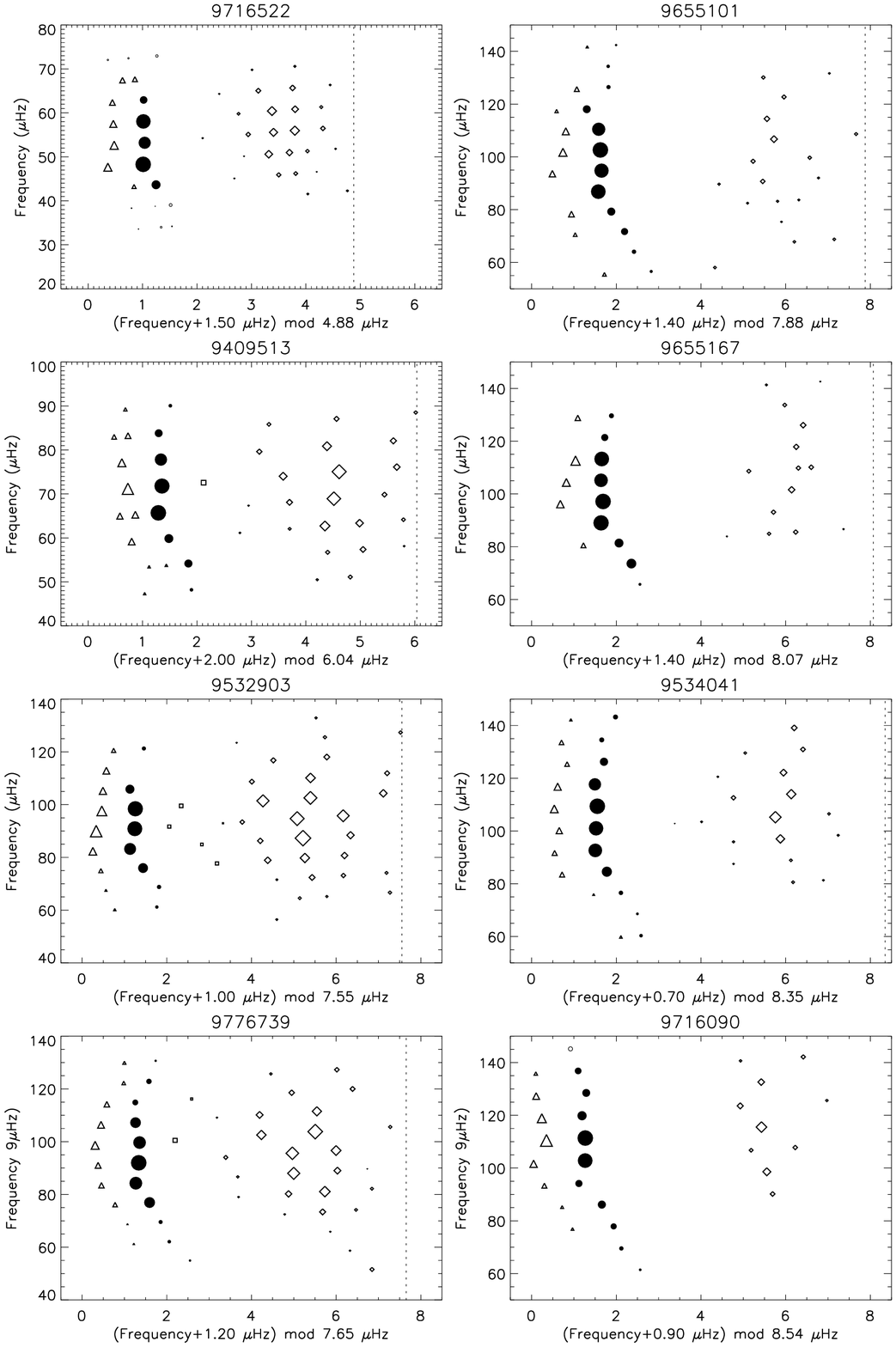}
 \caption{{\'E}chelle diagrams for all eight stars. The four left-most stars are those with lowest $\Delta\nu$ and richest mode structure. The modes are identified in the same way as in 
Fig.~\ref{fig.echelle} and the symbols are again scaled to the highest peak in each 
individual power spectrum. The ridges have been shifted along the x-axis using the constant given in the x-axis titles, for easier visual comparison between the stars. Several stars show similar 
ridge structures (curvature) which will be further investigated in a forthcoming paper. Some of the $l=2$ modes for KIC9409513, and possibly in KIC9716522, appear to be split in two. We assume
that this is due to the period spacing for $l=2$, which is expted to be smaller than the period spacing for $l=1$ by a factor of $\sqrt3$ \citep{jcd11}. A couple of modes with uncertain mode-identification
are marked as open circles for KIC9716522 and KIC9716090.}
\label{fig.allechelle}
\end{figure*}

The {\'e}chelle diagrams are shown for all eight stars in Fig.~\ref{fig.allechelle}. The four leftmost panels plot the {\'e}chelle diagrams of the stars with rich oscillation spectra, while the rightmost 
panels show the four stars with less pronounced $l=1$ modes. The differences already noted in the power spectra are seen from the {\'e}chelle diagrams as well. 
There is a wealth of information present in the oscillation spectra and {\'e}chelle diagrams for these stars, however the detailed asteroseismic analysis (including amplitudes, line widths, 
mode lifetimes, comparison of individual frequencies with model frequencies etc.) will be presented in a forthcoming paper, where the results for
the eclipsing binary stars in the cluster will aid the analysis by pinpointing the cluster isochrone and place additional constraints on the theoretical stellar models.

\section{Period spacings}

The mixed oscillation modes in evolved stars are characterized by a near equidistance in period \citep{jcd11}. The so-called observed period spacing $\Delta$P$_{\rm obs}$, which is some average 
of the period differences between consecutive pairs of modes, differ from the period spacing of the pure $g$-modes, $\Delta$P. However, in some
cases the latter can be inferred from the observed value 
\citep{2011Nature..471..608,2011AA...532..A86,2012ApJ..757..190}. \citet{2011Nature..471..608} showed that the observed period spacing can be used for distinguishing between hydrogen-shell-burning 
giants with $\Delta$P$_{\rm obs}$ values of $\sim$50 seconds and helium-core-burning giants with $\Delta$P$_{\rm obs}$ values of $\sim$100 -- 300 seconds and that 
the period spacings depend on both mass and evolution.

We have determined $\Delta$P$_{\rm obs}$ for our eight giants, again using an analysis largely based on \citet{mosumgaard14}. In the left panel of Fig.~\ref{fig.echelle} we show, as already discussed, 
the echelle diagramme 
for one of the stars (KIC9532903), separating the $l=0,2$ modes from the mixed $l=1$ modes. In the rightmost panel of Fig.~\ref{fig.echelle}, we show a histogram of the period differences 
between neighboring $l=1$ modes.
The x-axis plots the period spacing of neighboring modes $\left( \left|  P_i - P_j \right| \right)$ while the y-axis plots a measure of the power (amplitude) of the modes represented in each bin. 
If a pair of modes ($P_i$,$P_j$) have mode heights in the smoothed power spectrum of ($H_i$,$H_j$), then that pair contributes a value of $\sqrt(H_i \cdot H_j)$ to the sum of the power in the
corresponding bin. In this way, the analysis will be dominated by pairs where both modes have high signal-to-noise, and the region in the power spectrum where the oscillation power is highest is also
weighted highest. The histogram is then smoothed, and a Gaussian fit is used for determining the value of the average period difference, $\Delta$P$_{\rm obs}$. 
This is a different approach than the one used by \citet{2013ApJ...765L..41S}, however we have checked that the differences in the obtained values are only marginal. 

The $\Delta$P$_{\rm obs}$-values are listed in Table~\ref{tab.1}. They are in the range of $\sim$100 -- 150 seconds, which implies that these red giants are burning helium in 
their cores \citep{2011Nature..471..608}.

\begin{figure*}
\center \includegraphics[width=120mm]{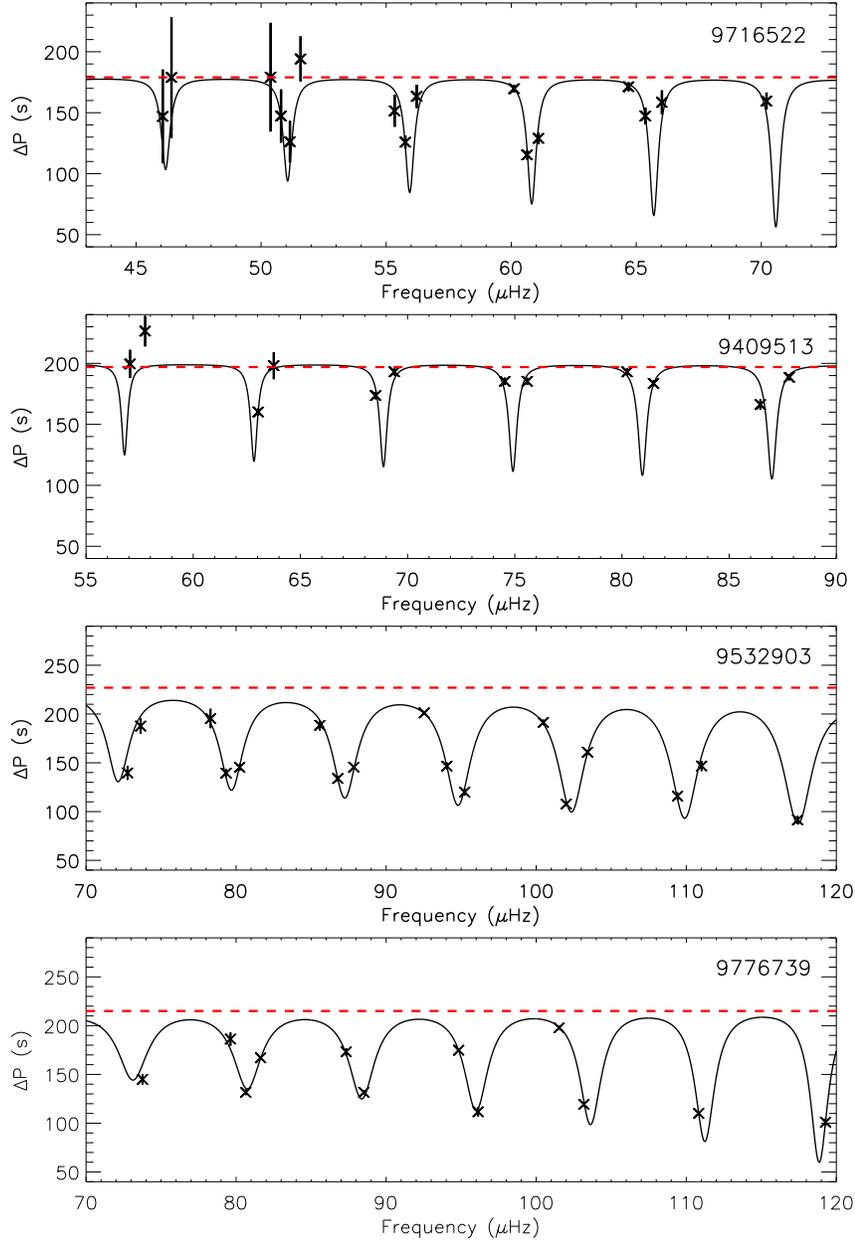}
\caption{Determination of the asymptotic period spacing, $\Delta$P, for four red giants in NGC\,6811. The plot shows period differences as a function of frequency; crosses represent period differences 
between neighboring $l=1$ modes, for period differences falling in the range between 50 and 250\,s (see the rightmost panel in Fig.~\ref{fig.echelle}). The solid line is the best (weighted) fit of a 
model consisting 
of a number of a Lorentzians whose width and depth were allowed to vary with frequency, subtracted from a constant value which is the asymptotic period difference $\Delta$P, see text 
for details. The asymptotic period differences are determined from the fits to be (from top to bottom) 179$\pm$4\,s, 200$\pm$5\,s, 227$\pm$6\,s and 215$\pm$3\,s, shown as the red-dashed lines 
in each panel.}
\label{fig.truedp}
\end{figure*}

Four of the stars have a sufficiently high number of detectable, mixed $l=1$ modes to allow a determination of the asymptotic period spacing $\Delta$P \citep{jcd11, 2012ASPC..462..200S}.
We used an empirical method described by \citet{2012ASPC..462..200S} to determine $\Delta$P; Fig.~\ref{fig.truedp} plots the individual period spacings versus the mean value of the two 
frequencies which gives the period spacing, for the four stars for which $\Delta$P is determined. Following \citet{2012ASPC..462..200S}, the data are fitted using the asymptotic period spacing of the 
$g$-modes, $\Delta$P, and a series of Lorentzian profiles separated in frequency by $\Delta\nu$, and for which the width and depth, and the changes with frequency in the width and depth, are fitted as 
free parameters. The fits are overplotted in Fig.~\ref{fig.truedp} and the asymptotic period spacings listed in Table~\ref{tab.1}. We tested the robustness of these fits, for example by varying 
initial guesses and omitting data points, and found the fits to be very stable. We compare the derived period spacings to theoretical models below. 

For the remaining four stars, we were not able to obtain stable fits to the individual period spacings versus frequency, due fewer frequencies and much lower oscillation amplitudes of especially 
the $l=1$ modes. 
For these stars, we set lower limits on the asymptotic period spacings based on the few individual period spacings we deemed trustworthy (i.e., where the individual frequency peaks in the power 
spectra have relatively high amplitude). These limits are listed in Table~\ref{tab.1} as well. 

We note that \citet{2016A&A...588A..82B} compared the method of \citet{2012ASPC..462..200S} for deriving $\Delta$P to the method of \citet{2012A&A...540A..143M} which relies on the asymptotic relation.
The conclusion of \citet{2016A&A...588A..82B} is that the two methods provide very similar results. Our measurements are furthermore in good agreement with the predictions for $\Delta$P$_{\rm obs}$ 
and $\Delta$P in \citet{2013ApJ..766..118}, their Fig.~1 (lower panel with overshoot, values in the range $\sim100-150$\,s for $\Delta$P$_{\rm obs}$ and $\sim160-230$\,s for $\Delta$P), 
except for KIC9409513, where our measured $\Delta$P$_{\rm obs}$ seems higher than expected. Finally, We remark that the fitted profiles for the two most evolved stars are different from the 
profiles of the two least evolved stars, as the more evolved stars have more narrow minima. This is related to the properties of the mixed modes as the stars evolve, however detailed studies 
of this effect requires 
individual modeling of the stars in question and will therefore be done at a later stage, when we can include constraints from the binary stars in the cluster.

\section{Stellar masses}

From the asteroseismic parameters, we can estimate the masses and radii of the eight red giants by using the relations between the asteroseismic parameters and stellar mass and
radius, see e.g. \citet{2012MNRAS..419..2077}:
\begin{small}
\begin{equation}
\frac{M}{M_{\odot}} \simeq \left( \frac{\nu_{\rm max}}{\nu_{\rm max,\odot}} \right)^{3} \left( \frac{\Delta \nu}{\Delta \nu} \right)^{-4} \left(\frac{T_{\rm eff}}{T_{\rm eff,\odot}} \right)^{3/2},
\end{equation}

\begin{equation}
\frac{R}{R_{\odot}} \simeq \left( \frac{\nu_{\rm max}}{\nu_{\rm max,\odot}} \right) \left( \frac{\Delta \nu}{\Delta \nu} \right)^{-2} \left(\frac{T_{\rm eff}}{T_{\rm eff,\odot}} \right)^{1/2}.
\end{equation}
\end{small}

For the solar values, we used $\nu_{\rm max,\odot} = 3090\,\mu$Hz, $\Delta \nu_{\odot} = 135.1\,\mu$Hz \citep{2011ApJ..743..143} and $T_{\rm eff,\odot} = 5777$\,K. The effective temperatures listed 
in Table~\ref{tab.1} are, for five of the stars, derived using the so-called {\sc SME}-method in \citet{2014MNRAS.445.2446M}. Since all stars have not been subject to a common spectroscopic study, we made use 
of other literature results for subsamples of the stars to put all stars on a common temperature scale. Specifically, KIC9534041 and 9716522 were analysed in \citet{2013MNRAS.434.1422M} together 
with three stars from \citet{2014MNRAS.445.2446M}. The mean $T_{\rm eff}$ difference of the three latter stars between the two studies (34\,K) was used to adjust the $T_{\rm eff}$ of KIC9534041 and 
9716522 to our adopted temperature scale, resulting in effective temperatures of 5027 and 4826\,K, respectively. For KIC9409513, we found its effective temperature along with that of KIC9695101 in the 
APOKASC catalog \citep{2014ApJS..215...19P}, and used the temperature difference for KIC9695101 to place KIC9409513 on our scale.  Although it is desirable to measure the effective temperatures of all 
the stars in a homogeneous way for later detailed studies, the temperatures obtained using the corrections described here are sufficient for our current analysis. The results for the masses and radii 
are listed in Table~\ref{tab.1}, indicating masses for the red giants of just above 2\,M$_{\odot}$. The uncertainties on mass and radius are obtained using propagation of errors for $\nu_{\rm max}$, $\Delta \nu$, 
and $T_{\rm eff}$ where for the latter we adopted an uncertainty of 100\,K in agreement with \citet{2014MNRAS.445.2446M}.

The scaling relations provide a model-indepen-\\dent estimate of stellar properties that can be refined from our knowledge of stellar evolution. With this in mind, we fit the observed asteroseismic 
and atmospheric quantities to a grid of BaSTI \citep{basti04} isochrones using the BAyesian STellar Algorithm \citep[BASTA,][]{vsa15, vsa17}. We take into account deviations in the scaling relations as 
a function of metallicity, effective temperature, and evolutionary phase using the corrections from Serenelli et. al.~2017 (in preparation), and determine the stellar properties given in Table~\ref{tab.3}. 
These results indicate a mass-sequence among the eight stars, where lower $\nu_{\rm max}, \Delta\nu$ means higher mass, and that the masses are slightly higher than 2M$_{\odot}$, in agreement with 
those derived from the scaling relations above. In this we have assumed a common age for all eight stars of $1.0\pm0.1$\,Gyr and solar metallicity
following the results of \citet{2014MNRAS.445.2446M}. Effective temperature offsets could influence the absolute values of the masses obtained by asteroseismology, but not the basic results of a 
mass sequence in the cluster. We have done tests of our obtained masses by increasing/decreasing the $T_{\rm eff}$ values by 50\,K, and found variations no larger than 0.03\,M$_{\odot}$ in the individual 
stars and the persistence of a mass sequence as a function of evolution. We note that the 
formal model uncertainty estimates in Table~\ref{tab.3} become unrealistically low for the bottom couple of stars, however as we only use the models for corroborating the estimated masses, 
this is not important for the present paper.

\section{Comparison with theoretical models}
The stellar properties determined in the previous section are based on the acoustic global seismic parameters $\nu_{\rm max}$ and $\Delta \nu$, which are sensitive to the ratio of the stellar 
mass and radius. More detailed information about the stellar core can be obtained from the analysis of gravity modes that penetrate deep into the stellar interior.
\citet{2013ApJ..766..118} studied the correlations between period spacings in red giants and the characteristics of the helium core, based on theoretical models. The stellar masses derived above imply that the red giants in NGC\,6811 have masses close to the so-called transition mass, separating stars into those that go through a helium flash and those that ignite helium under non-degenerate conditions. 
This means that these stars have relatively low helium-core masses and belong to the secondary clump \citep{1999MNRAS.308..818G}. This conclusion is corroborated by our four derived asymptotic period 
spacings in the range 180 -- 220\,s, which is in agreement with the values expected for masses close to the transition mass, see Fig.~1 in \citet{2013ApJ..766..118}.

Period spacings of red giants have the potential of constraining convective-core overshoot in theoretical stellar models \citep{2011Nature..471..608,2013ApJ..766..118}.
To have a first look at this for NGC\,6811, we calculated sequences of models of evolved stars in order to compare our derived values of $\nu_{\rm max}$, $\Delta \nu$ and $\Delta$P 
with those of the models, in a similar procedure as used by \citet{hjg17}. The models were computed with the 
Garching Stellar Evolution Code (GARSTEC, Weiss \& Schlattl 2008) at solar metallicity using the \citet{1998SSRV..85..161G} mixture, 
the OPAL equation of state \citep{1996ApJ..456..902R, 2002ApJ..576..1064R}, OPAL opacities for high temperatures \citep{1996ApJ..464..943I} and those of \citet{2005ApJ..623..585F} for low temperatures, the 
NACRE compilation of nuclear reaction rates \citep{1999NuPhA.656....3A} including the updated $^{14}N(p,\gamma)^{15}O$ reaction from \citet{2004PhLB..591...61F}, and the mixing-length theory of 
convection as described in \citet{2012sse..book.....K}. Overshoot is implemented as a diffusive process using an exponential decay of the convective velocities based on 2D simulations of stellar atmospheres \citep{freytag96}, with the diffusion constant given by:
 
\begin{equation}
\centering
D_\mathrm{ov}(z)=D_0 \exp\frac{-2z}{\xi H_\mathrm{p}}\,.
\label{eq:dnuscal}
\end{equation}

Here $z$ is the distance to the convective border, $D_\mathrm{0}$ is a constant derived from the mixing-length theory convective velocities, $H_\mathrm{p}$ is the pressure scale height and $\xi$ is 
an efficiency parameter. The code includes a geometrical cut-off factor to limit the amount of overshoot if convective cores are small and is calibrated to prevent the survival of the 
pre-main-sequence convective core appearing from non-equilibrium $^3$He burning in a solar model. Under this prescription the calibrated main-sequence overshoot 
efficiency from isochrone fits to open cluster turn-offs is $\xi=0.016$ \citep[see, e.g.,][for details]{2010ApJ...718.1378M,vsa11}, and it is the value we used for main-sequence overshoot in 
all the models where this effect is taken into account. However, in the present analysis we removed the geometrical cut-off during the helium-burning phase to allow a higher degree of control in 
the amount of mixing beyond the formal convective boundary to reproduce the measured period spacing values (see below). For this reason the efficiency values quoted for the core helium-burning 
phase are not directly comparable to the $\xi=0.016$ efficiency applied in the main sequence. The asymptotic period spacings for selected models in the helium-core-burning phase were determined 
from the integral of the Brunt-V{\"a}is{\"a}l{\"a} frequency, while estimates of the large frequency separation and frequency of maximum power were obtained from the scaling relations.

\begin{figure*}
\center \includegraphics[width=110mm]{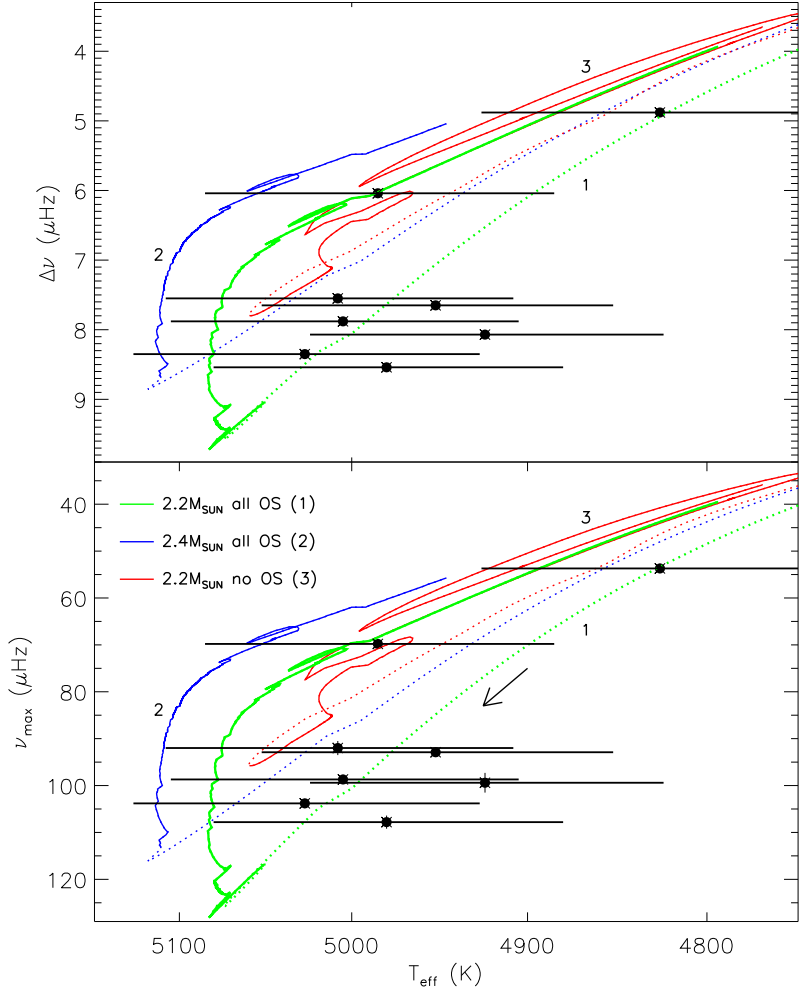}
\caption{Comparison between our measured values of $\Delta\nu$ and $\nu_{\rm max}$ and theoretical stellar models with different masses and convective-core overshoot characteristics. The green and blue
tracks differ only in mass, having convective-core overshoot in both the main-sequence and helium-burning phases, while the red track does not include any overshoot. The tracks are
identified using colors and labels 1--3; the red and green curves coincides in parts of the diagramme which causes the red curve to appear broken in certain parts. 
The dotted lines indicate fast evolution before the stars settle in the clump. A star evolves from the right hand side towards the lower left, as indicated by the arrow in the bottom panel, 
and then up and back towards the upper right hand side. The model without overshoot does not match our observed values for $\Delta\nu$ and $\nu_{\rm max}$ as it does not predict sufficiently 
high values for these two parameters. The models show saw tooth behaviour which is a consequence of small changes in the size of the convective core during the He-burning phase. When using the 
Schwarzschild criterion for defining the convective boundaries, small variations in the opacity outside the convectively unstable region extend the size of the convective core. 
This is accompanied by small changes in the effective temperature and luminosity which in turn affect the computation of $\Delta\nu$ and $\nu_{\rm max}$ from the scaling relations (see Eqs. 5 and 6).
}
\label{fig.models1}
\end{figure*}

\begin{figure*}
\center \includegraphics[width=120mm]{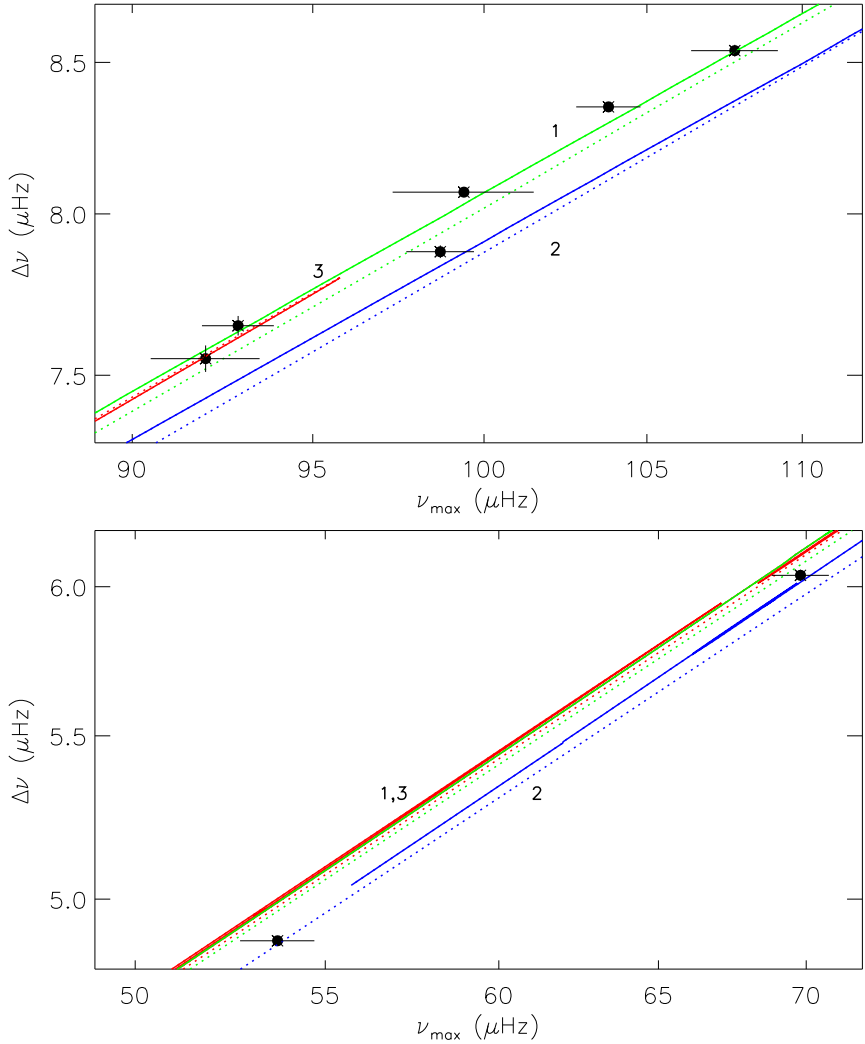}
\caption{Plots showing $\nu_{\rm max}$ and $\Delta\nu$ (corresponding to Eq.~4), in the region of the two most evolved stars (bottom panel) and in the region of the six stars with similar parameters 
($\nu_{\rm max}$ above 90\,$\mu$Hz, upper panel). The tracks are identified using the same colors and labels $1-3$ as in Fig.~\ref{fig.models1}. The green and the red curves coincide in the lower panel.
In general, our measurements matches the theoretical track with a mass of 2.2\,M$_{\odot}$ (green curve) in the upper panel, and the 2.4\,M$_{\odot}$ track (blue curve) in the lower panel, in agreement
with the expectation of a higher mass for the more evolved stars. Such a conclusion is however uncertain due to the errorbars on $\nu_{\rm max}$.
}
\label{fig.models2}
\end{figure*}

We have calculated models of 2.2\,M$_{\odot}$ stars, based on the masses derived from the scaling relations above, and 2.4\,M$_{\odot}$ models for comparison. 
The models have been computed 
with 1) overshoot both on the main sequence ($\xi=0.016$) and in the core-helium-burning phase (see above for details), 2) with overshoot only in either the main-sequence ($\xi=0.016$) or in the 
helium-burning phase ($\xi$ variable), and 3) without any overshoot at all. In Fig.~\ref{fig.models1} and~\ref{fig.models2}
we compare three of the models to the values of $\Delta\nu$ and $\nu_{\rm max}$ for our eight stars. 
Two of the models (green and blue) include convective overshoot on both the main sequence and in the core-helium-burning phase but differ in mass, while the last model (red) has the same mass as the green, 
but has no overshoot. A model with overshoot only in the core-helium-burning phase is very similar to the model without any overshoot, just as a model with 
overshoot in the main sequence only is very similar to the model with overshoot in both phases. These models are therefore not shown in Fig.~\ref{fig.models1} and~\ref{fig.models2}.

These plots indicate that the 2.2\,M$_{\odot}$ model with overshoot matches the data better than the corresponding 2.4\,M$_{\odot}$ model for the least evolved stars, 
and that, assuming that a mass of 2.2\,M$_{\odot}$ is indeed representative for these stars, a model without overshoot, or with overshoot in the core-helium-burning phase only, 
cannot reproduce the measured values of $\nu_{\rm max}$ and $\Delta\nu$. 
Convective-core overshoot on the main sequence (with an efficiency of the order of the calibrated $\xi=0.016$ value) therefore seems necessary for matching the observed asteroseismic parameters 
of $\nu_{\rm max}$ and $\Delta\nu$ for the secondary-clump stars in NGC\,6811. However, we can not constrain overshoot in the core-helium-burning phase because $\nu_{\rm max}$ and $\Delta\nu$
are not sufficient in this regard. 
We note that it is possible that a sequence of models with slightly different masses and overshoot may also match our measurements. It is, however, beyond the scope of the present paper to 
explore this any further.

\begin{figure*}
\center \includegraphics[width=150mm]{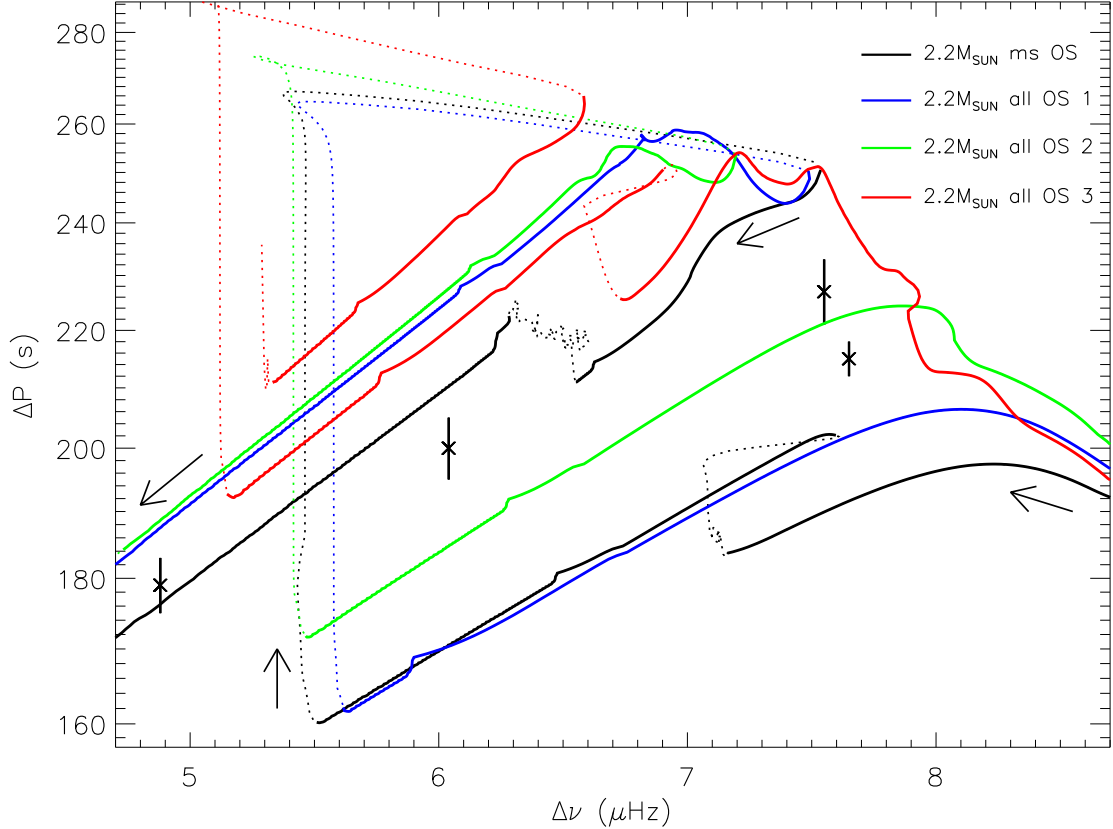}
\caption{Comparison between our measured values of $\Delta\nu$ and $\Delta$P and theoretical models in the core-helium-burning phase. The black model includes overshoot on the main sequence only 
while the remaining three models include various degrees of overshoot in the core-helium-burning phase as well. As indicated by the arrows, the evolution starts in the lower right-hand corner; 
the tracks generally evolves towards the left and ultimately leaves the diagram on the left-hand side, except for the red track which has not been calculated all the way through the core-helium-burning 
phase. During the evolution, the tracks undergo some sudden changes in the core with sharp increases in the period spacings and reversed evolution in $\Delta\nu$ as a result, followed by evolution 
back towards lower $\Delta\nu$ again. These are the so-called breathing pulses discussed in the text; they are shown as dotted lines in the diagram to indicate that they are short phases of the evolution.
The wiggles in the first part of the red curve is as in Fig.~\ref{fig.models1} due to the changing size of the convective core. The red curve has the highest efficiency of overshooting, which 
produces the largest mix beyond the formal core boundary and results in stronger extension and contraction of the convectively unstable core. The changes in the size of the convective region impact 
directly the calculation of the asymptotic period spacing due to variations in the Brunt-V{\"a}is{\"a}l{\"a} frequency.
} 
\label{fig.sequence}
\end{figure*}

In Fig.~\ref{fig.sequence} we compare the 
measured values of ($\Delta\nu$,$\Delta$P) with theoretical models with different amounts of convective-core overshoot in the helium-core-burning phase. The black model has overshoot on the 
main sequence only, using the calibrated $\xi=0.016$ value, while the other models (labeled {\it all OS\,1, OS\,2} and {\it OS\,3}) also have increasing amounts of overshoot in the helium-burning 
phase, resulting in different sizes of the convective cores. 
The figure shows that models with overshoot on the main sequence only does not reproduce the measured combinations of $\Delta\nu$ and $\Delta$P, at least not in all parts of the diagram. 
The models can be tweaked by adding overshoot in the core-helium-burning phases, which have been done for three additional models also included in the figure; 
the green model (labeled {\it all OS 2}) fits two of the data points reasonably well, whereas the black model 
seems to be the better fit for the two more evolved stars with lowest $\Delta\nu$. We would not expect, however, to find a single evolutionary track which matches all the data points because we are 
in fact observing a sequence of masses, as mentioned above. Despite this, we haven't yet found a scenario which explains all the measurements.
Further refinement is beyond the scope of the the present paper, as the eclipsing binaries will provide further 
constraints for the stellar models, however Fig.~\ref{fig.sequence} demonstrates the potential of red-giant period-spacings for constraining the amount of mixing beyond the core in the helium-burning 
phase as well as in the main sequence. 

The models presented here display so-called {\it  breathing pulses} which are sudden changes in evolution due to instabilities at the core, resulting in mixing of 
additional helium into the core and hence increasing the time spent in the helium-burning phase slightly. It is still a matter of debate whether these instabilities actually occur in stars, 
see e.g. \citet{2005astro.ph..6161C}; Fig.~\ref{fig.sequence} suggests that further investigation, using measured ($\Delta\nu$,$\Delta$P)-values and dedicated modeling, will provide input to this 
discussion. The breathing pulses seen in our Fig.~\ref{fig.sequence} seems very large and have a significant effect on the evolutionary tracks at low $\Delta\nu$; \citet{2015MNRAS..452..123}
show a diagram similar to ours (their Fig.~19), however their models are not shown down to sufficiently low $\Delta\nu$, and we therefore cannot judge if their models display 
breathing pulses or not and how their models compare with our measurements at low $\Delta\nu$. For their model with maximal overshoot, it seems unlikely that they would be able match our data points 
as the predicted period spacings are too high.

We finally note that according to our measured period spacings, the two most evolved stars (with lowest $\nu_{\rm max}$ and $\Delta\nu$) are still
burning helium in their cores, as AGB stars will have much lower period spacings, similar to those of the RGB stars. \citet{2013ApJ...765L..41S} describes how the $\Delta$P-values will 
very quickly (within 1 Myr) drop at the end of core-helium burning, which is also found by \citet{2015MNRAS..452..123}, who discuss how stars leaving the secondary clump to become AGB stars are in fact 
still burning helium in their cores; this is supported by our 2.2\,M$_{\odot}$ model which indicate that the most evolved of these two stars still has about 10 \% helium left in its core.

\section{Suppressed modes}

As is seen in Fig.~\ref{fig.power}, the power spectra of our eight stars split them in two groups with the four most evolved stars (lowest $\nu_{\rm max}$) showing rich spectra while the other 
four show much fewer $l=1$ modes, which is the reason we were not able to derive the asymptotic period spacings for the latter. The stellar parameters for these 
stars are very similar, especially for the six stars with $\Delta\nu$
values above 7\,$\mu$Hz, rendering an explanation of the observed differences based only on structural differences unlikely. It is therefore reasonable to look for other explanations and,   
as mentioned in the introduction, mode suppression of $l=1$ and higher order modes have been suggested to arise due to strong internal magnetic fields.

\begin{figure*}
\center \includegraphics[width=118mm]{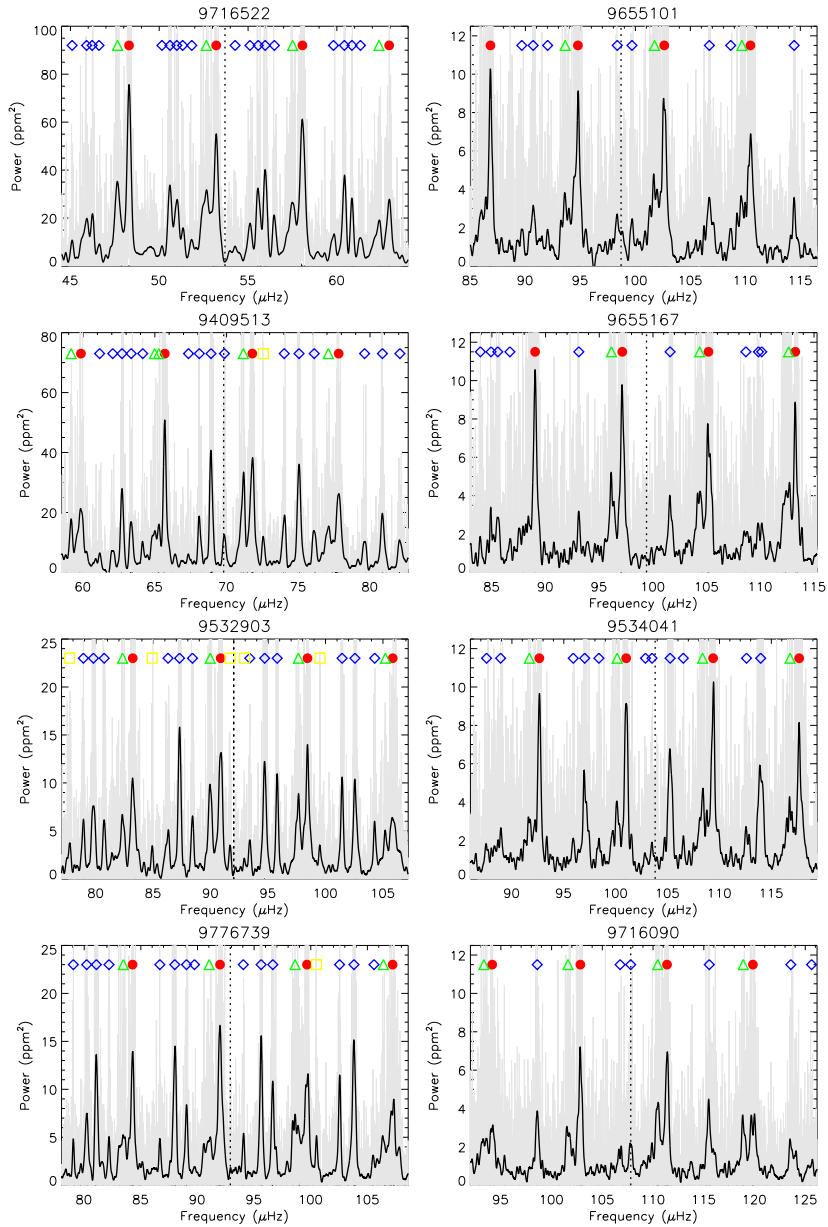}
 \caption{The central regions of the power spectra with the stellar backgrounds subtracted for the eight stars, arranged in the same order and having modes identified using the same symbols 
as in Fig.~\ref{fig.allechelle}. The color-coding of the symbols corresponds to the color-coding in Fig.~\ref{fig.collapse}: filled, red circles are $l=0$, blue diamonds are $l=2$, green 
triangles are $l=2$ and yellow squares are $l=3$. The original, background-corrected spectra are shown in light-gray, while the solid black lines plot smoothed versions of those spectra. These are 
the spectral regions used for the folded spectra in Fig.~\ref{fig.collapse}. The vertical dotted lines indicate $\nu_{\rm max}$. The y-scales in the right-hand plots are half of those of the two bottow 
panels on the left-hand side.
}
\label{fig.central}
\end{figure*}

\begin{figure*}
\center \includegraphics[width=118mm]{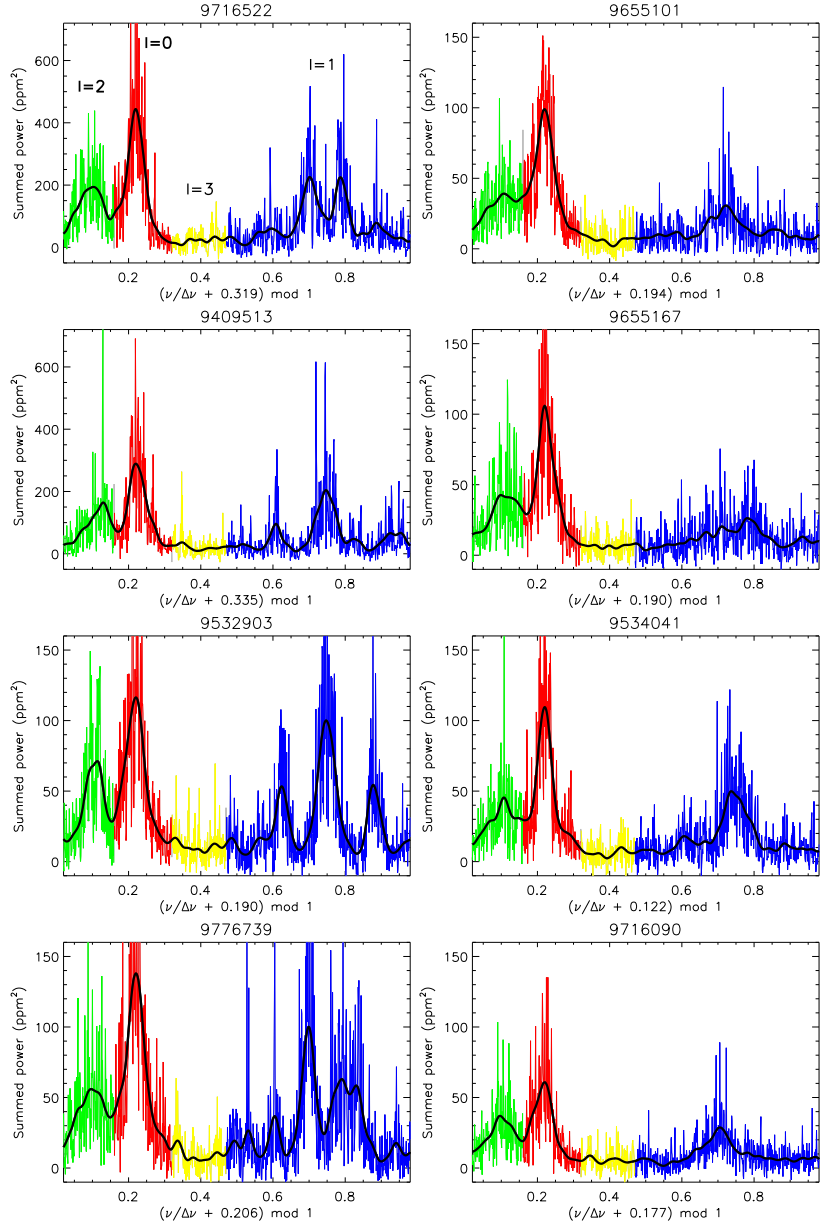}
 \caption{The central regions of the background subtracted power spectra of the eight stars folded with their large frequency separations, and arranged in the same order as in 
Fig.~\ref{fig.allechelle}. Shifts along the x-axis have been introduced in order to allign the regions corresponding to the different $l$ values. 
The data shown in this figure is used for measuring squared visibilities for the $l=1$ and $l=2$ modes. Notice that the two upper leftmost panels have y-scales different from the other 
6 panels. The different colors correspond to regions of different $l$-values and the solid black lines are smoothed versions of the folded power spectra.
}
\label{fig.collapse}
\end{figure*}

Before we discuss this in further detail, we investigate whether our four stars with highest $\Delta\nu$ actually do show suppressed modes. This is done by measuring the squared mode 
visibilities \citep{2012A&A...537A..30M}, following the methods described in \citet{2016Natur.529..364S,2016PASA...33...11S}. For this, we used versions of the power spectra where the background signal 
described in Sect.~3 had been subtracted. These power spectra showed minor, positive offsets from zero in the regions of the oscillations; in order to correct for this, we identified the regions of the 
four central orders of the $l=3$ modes, which have very small oscillation amplitudes, and used for each star the median value of these regions to correct the zeropoints.
For each of the eight stars, we extracted a region of the power spectrum containing the four central $l=0$ peaks, 
see Fig.~\ref{fig.central}. These regions were then folded using the large frequency separations, resulting in the diagrams shown in 
Fig.~\ref{fig.collapse}. Again following \citet{2016Natur.529..364S,2016PASA...33...11S}, the folded diagrams were divided in regions associated with $l=0-3$, and the summed power in each region were 
calculated. The mode visibilities of the non-radial, $l=1,2$ modes were calculated as the summed power in the region corresponding to the given $l$-value, divided by the summed power in the region of $l=0$. 
The results are given in Table~\ref{tab.4} for $l=1$ and $l=2$ (we have not done this for the few suspected $l=3$ modes).
Comparing the visibilities of $l=1$ and $l=2$ we see differences between the stars for $l=1$ (first data column in Table~\ref{tab.4}), while this is not the case for $l=2$. The values of the mode 
visibilities for $l=1$ and $l=2$ (first and third data column in Table~\ref{tab.4}) can be directly compared with the theoretical predictions in \citet{2011A&A...531A.124B}. 
These authors predict $l=1$ visibilities
for non-suppressed stars of 1.54, and $l=2$ visibilities of 0.58. For two of our stars, KIC9532903 and KIC9776739, the visibilities for $l=1$ mathces the theoretical predictions, which is not the case
for the remaining six stars. For $l=2$ all but one of our stars matches reasonably well with the theoretical predictions. For one star, KIC9716909, the squared visibility for $l=2$ is higher than for 
the remaining stars, however we assume that this is due to a lack of power in $l=0$, see Fig.~\ref{fig.collapse}.

For comparison, \citet{2016arXiv160203056C} published predictions for magnetic suppression of $l=1$ and $l=2$ modes for core-helium burning giants. According to their Fig. 1, the squared
visibilities relative to those of stars with no mode suppression should be in the range $0.05-0.20$ for $l=1$ and $0.40-0.75$ for $l=2$ for core-helium burning stars with masses between
2.0 and 2.5\,$M_{\odot}$ and radial mode lifetimes of 10-30 days.
In Fig.~\ref{fig.central} we show the smoothed and un-smoothed power spectra for each of the eight stars. We have fitted lorentzian profiles to a number of isolated modes in each of the eight stars
(not shown) and find that the mode lifetimes are in the range of 10-30 days (see Table~\ref{tab.1}), and our results are therefore directly comparable to the predictions.

To compare with the result of \citet{2016arXiv160203056C}, we have in data column two in Table~\ref{tab.4} divided the measured visibilities with the mean value
for $l=1$ for KIC9532903 and KIC9776739. Although we may see mode suppression of $l=1$ for some of the stars, it is much less than predicted by the magnetic suppression theory, and for $l=2$ we do not
see any suppression at all. This allows for two different interpretations of our measurements. 

Previous investigations of $l=1$ visibilities for red-giant stars observed with {\it Kepler} show that the stars separate into two branches \citep{2016Natur.529..364S,2011AA...532..A86} 
Stars on the branch with low visibilities are referred to as having suppressed modes. However, the scatter among stars on the branch corresponding to unsuppressed modes is much larger than predicted 
by theory due to differences in stellar parameters \citep{2011A&A...531A.124B}. Therefore, one interpretation of our measurements is that none of the stars show suppressed modes by the definition 
in the literature, but there are small but real star-to-star differences of unknown physical origin. 

Alternatively, the four least evolved stars might be showing suppressed modes of $l=1$, but much less than predicted by the current theory of magnetic suppression. This may be related to the issue
mentioned by \citet{NewMosser} that the theory assumes close to full suppression \citep{2015Sci...350..423F}; however we do observe mixed $l=1$ modes in all our stars (see Fig.~\ref{fig.allechelle} and
Fig.~\ref{fig.central}), which can only be the case if there is power in the $g$-modes.

In the following, we consider what conclusions might be reached by accepting the current theory of magnetic suppression for either of these two scenarios, starting with the interpretation of the four 
least evolved stars showing suppressed modes while the others do not. We return to the other interpretation later.

\citet{2016arXiv160203056C} make theoretical predictions for what the observations of suppressed modes (or not)
in helium-core-burning giants can tell us about the destruction of internal magnetic fields during the evolution of giant stars. The basic idea is that the magnetic field will likely be consumed in
convective regions in the star after the main sequence. Therefore, a magnetic field is only able to survive in regions of the star that does not become convective at any time after the main sequence.
According to the model calculations by \citet{2016arXiv160203056C}, their Fig. 6, there are no such regions once the star reaches the helium-burning phase if it is below the transition mass and goes
through the helium flash. But if the star is above the transition mass there will be a radiative region above the helium-burning convective core which is left untouched by convection where a magnetic
field could survive.

\begin{figure*}
\center \includegraphics[width=125mm]{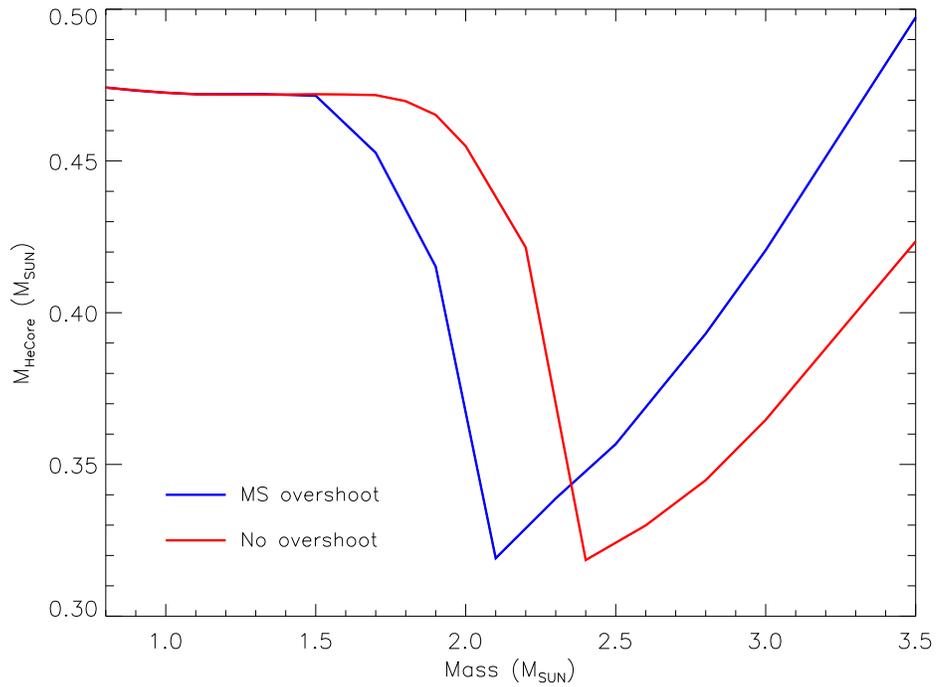}
\caption{Helium core-sizes as a function of stellar mass for the BaSTI models discussed in Sect.~6. The blue (leftmost) curve is for models where overshoot is included on the main sequence, while the 
red (rightmost) curve is for models without overshoot. With a transition mass of 2.1\,M$_{\odot}$ for the models with overshoot on the main sequence, the results in this paper (Tables~\ref{tab.1} and 
\ref{tab.3}) indicate that all eight stars have masses close to the transition mass. We thank S. Cassisi for providing the data for this figure.
}
\label{fig.basti}
\end{figure*}

In Fig.~\ref{fig.basti} we show the size of the helium core for the BaSTI-models introduced above. Two model sequences are shown, one
with overshoot on the main sequence, and one without. As we have shown, overshoot on the main sequence is necessary to match our observations, indicating that the stars discussed in this paper have
masses close to the transition mass in the models, which is about 2.1\,M$_{\odot}$ for the models with overshoot.

According to Fig.~\ref{fig.power} and Table~\ref{tab.4}, we see suppressed modes for the lowest-mass stars and no mode suppression for the higher-mass stars. Therefore, within the theory of
magnetic suppression, the difference between the two groups is not related to the transition mass, since that would create the opposite scenario. Instead, our observations of four stars
with suppressed modes in their early stages of helium core-burning provides
evidence that they are indeed above the transition mass. By extension, the four more evolved stars are also above the transition mass as their later evolutionary stages dictate that they reached
helium core-burning earlier, which happened because they were more massive.

To explain why the four most evolved stars do not show mode suppression we need to adopt a more speculative part of the analysis of \citet{2016arXiv160203056C}. The suggestion is that the consumption
by the convective core of the part of the magnetic field inside the core, will also destabilize and thereby remove the magnetic field from the radiative regions above the convective core.
If this process takes
some time, then it can explain why the four most evolved stars in our sample do not show suppressed modes while the four least evolved ones do; the convective helium-burning core destroys the magnetic
field, but not immediately.
Alternatively, since our model comparisons in Fig.~\ref{fig.sequence} suggest that the transition between suppressed and non-suppressed modes happens at a mass close to where the models show
the maximum extension of the core,
it could be that at this point the convective core has grown big enough to eliminate the last regions that has remained radiative since the main sequence, see Fig. 6 of \citet{2016arXiv160203056C}.

The fact that the stars fall in two groups with very different oscillation characteristics may be by chance, as mode suppression is only seen in a subset of the red giants.
The masses of our targets are similar to the ones showing close to 50\% occurrence rates for suppressed modes among RGB stars \citep{2016Natur.529..364S}.
Assuming there is no evolution effect (no change since the RGB) and a 50\% occurrence rate for suppressed modes, there would be a $\frac{1}{2^8}=0.4\%$ probability of reaching our observed 
grouping by chance.

Returning to the alternative interpretation of no mode suppression, it seems from \citet{2016Natur.529..364S} that the intrinsic scatter among the stars whith no suppression is comparable to the 
differences we see between our stars. Therefore, we are perhaps just seeing stars at different sides of a population showing a range of visibilities.  
Our sample is much more homogeneous than 
the sample in \citet{2016Natur.529..364S} as our cluster stars share common properties such as chemical composition and age, but as already noted by \citet{2011AA...532..A86} the observed variation of 
visibilities is much larger than predicted due to differences in stellar parameters \citep{2011A&A...531A.124B}. Thus, the physical cause for the different visibilities is currently unknown.

According to Stello et al. (2016), around 50\% of giant stars in our mass range show suppressed modes on the RGB. If none of our eight stars, which are all in the later secondary clump phase, 
show magnetically suppressed modes, this suggests that the magnetic fields in red-giant stars disappears between the RGB and secondary clump phases. If not, there is only a $\frac{1}{2^8}=0.4\%$ probability 
of finding no star with mode suppression in our sample.

Regardless of which scenario turns out to be the correct, the eight stars presented in this paper constitutes an interesting sample for investigating mode suppression.

\section{Conclusions and outlook} 
We have analyzed data from {\it Kepler} for eight red-giant stars in the open cluster NGC\,6811. From the asteroseismic analysis, we find these eight stars to be helium-core-burning, secondary-clump stars. 
Two stars are on the verge of evolving into the AGB-phase. Comparison with stellar models indicate that these stars can be used for constraining the stellar models, including the amount of convective-core
overshoot in the main-sequence and helium-core-burning phases. The possible division of the eight stars in two groups with different mode visibilities points to some 
evolutionary effect being at play. This may be mode suppression by internal magnetic fields, as discussed in recent literature, however we observe a lower amount of suppression than currently predicted 
by theory.

As a next step, we will extend the theoretical analysis to include comparisons with the observed period spacings as well as with the individual oscillation frequencies (the {\'e}chelle diagrammes), and we 
will combine the asteroseismic constraints with those emerging from an ongoing analysis of eclipsing binary stars in NGC\,6811. 

\acknowledgments
Funding for the Stellar Astrophysics Centre is provided by The Danish National Research Foundation (Grant agreement no.: DNRF106). The research was supported by the ASTERISK project 
(ASTERoseismic Investigations with {\it SONG} and {\it Kepler}) funded by the European Research Council (Grant agreement no.: 267864). 
In loving memory of Nikolaj Holmbo Arentoft. These stars now sing for you.

\clearpage
\newpage

\begin{sidewaystable}
\caption{Asteroseismic parameters and other properties for the red giants in NGC\,6811. For further properties including membership information, see \citet{Sand2016..831..11S}, their Table~6.
}
\begin{tabular}{ccccccccccccc}
\hline
\noalign{\smallskip}
KIC     & $\Delta\nu$ ($\mu$Hz) & $\delta_{\rm 02}$ ($\mu$Hz) & $\epsilon$   & $\nu_{\rm max}$ ($\mu$Hz)  & $\Delta$P$_{\rm obs}$ (s) & $\Delta$P (s) & T$_{\rm eff}$ (K) & M$_{\rm scal}$(M$_{\odot}$) & R$_{\rm scal}$(R$_{\odot}$) & log\,$g_{\rm seis}$  & $\tau_{l=0}$ (d)  \\
\noalign{\smallskip} \hline
9716522 & 4.88$\pm$0.01  & 0.56$\pm$0.03  & 0.902$\pm$0.003   &  53.7$\pm$1.0 & 147$\pm$8 & 179$\pm$4      & 4826$\pm$100      & 2.33$\pm$0.17  &12.12$\pm$0.30   & 2.64   &  29$\pm$7  \\
9409513 & 6.04$\pm$0.02  & 0.63$\pm$0.02  & 0.887$\pm$0.005   &  69.8$\pm$1.0 & 190$\pm$3 & 197$\pm$5      & 4985$\pm$100      & 2.29$\pm$0.12  &10.46$\pm$0.20   & 2.76   &  29$\pm$13 \\ 
9532903 & 7.55$\pm$0.04  & 0.91$\pm$0.01  & 1.025$\pm$0.009   &  92.0$\pm$1.5 & 135$\pm$3 & 227$\pm$6      & 5008$\pm$100      & 2.15$\pm$0.12  & 8.83$\pm$0.17   & 2.88   &  20$\pm$3  \\ 
9776739 & 7.65$\pm$0.03  & 0.96$\pm$0.02  & 1.014$\pm$0.006   &  92.9$\pm$1.0 & 136$\pm$3 & 215$\pm$3      & 4952$\pm$100      & 2.07$\pm$0.15  & 8.64$\pm$0.21   & 2.88   &  20$\pm$4  \\
9655101 & 7.88$\pm$0.02  & 0.88$\pm$0.05  & 1.027$\pm$0.004   &  98.7$\pm$1.0 & 134$\pm$3 & $\gtrsim$170   & 5005$\pm$100      & 2.24$\pm$0.11  & 8.70$\pm$0.14   & 2.91   &  18$\pm$5  \\
9655167 & 8.07$\pm$0.01  & 0.82$\pm$0.02  & 1.032$\pm$0.003   &  99.4$\pm$2.1 & 104$\pm$8 & $\gtrsim$140   & 4924$\pm$100      & 2.03$\pm$0.15  & 8.29$\pm$0.20   & 2.91   &  23$\pm$8  \\
9534041 & 8.35$\pm$0.01  & 0.87$\pm$0.04  & 1.098$\pm$0.003   & 103.8$\pm$1.0 & 120$\pm$2 & $\gtrsim$155   & 5027$\pm$100      & 2.09$\pm$0.10  & 8.17$\pm$0.13   & 2.93   &  19$\pm$2  \\
9716090 & 8.54$\pm$0.02  & 0.92$\pm$0.05  & 1.041$\pm$0.005   & 107.8$\pm$1.4 & 100$\pm$3 & $\gtrsim$125   & 4980$\pm$100      & 2.10$\pm$0.13  & 8.07$\pm$0.17   & 2.94   &  14$\pm$5  \\
\hline
\noalign{\smallskip}
\end{tabular}
\label{tab.1}
\end{sidewaystable}

\clearpage

\begin{table}
\small
\caption{List of frequencies with undertainties, signal-to-noise ($SN$) and mode identification for KIC9532903.
The $l=1$-modes are mixed and no $n$-values are assigned.}
\begin{tabular}{ccc}
\hline
\noalign{\smallskip}
$\nu$ ($\mu$Hz)    & SN & Mode ID \\
\noalign{\smallskip} \hline
\vspace{1mm}
 56.459$\pm$0.051  &    1.7  &  $l=1$       \\
 60.184$\pm$0.045  &    2.2  &  $l=2$, $n=6$ \\
 61.177$\pm$0.038  &    3.1  &  $l=0$, $n=7$ \\
 64.553$\pm$0.039  &    2.9  &  $l=1$         \\
 65.193$\pm$0.045  &    2.2  &  $l=1$         \\
 66.683$\pm$0.038  &    3.2  &  $l=1$         \\
 67.527$\pm$0.046  &    2.1  &  $l=2$, $n=7$ \\
 68.781$\pm$0.034  &    4.2  &  $l=0$, $n=8$ \\
 71.564$\pm$0.048  &    2.0  &  $l=1$         \\
 72.396$\pm$0.030  &    6.1  &  $l=1$         \\
 73.135$\pm$0.033  &    4.6  &  $l=1$         \\
 74.153$\pm$0.038  &    3.1  &  $l=1$         \\
 74.959$\pm$0.033  &    4.4  &  $l=2$, $n=8$  \\
 75.955$\pm$0.027  &   10.4  &  $l=0$, $n=9$  \\
 77.701$\pm$0.037  &    3.4  &  $l=3$?        \\
 78.899$\pm$0.029  &    7.1  &  $l=1$         \\
 79.777$\pm$0.027  &   10.3  &  $l=1$         \\
 80.714$\pm$0.029  &    7.2  &  $l=1$         \\
 82.321$\pm$0.028  &    8.6  &  $l=2$, $n=9$  \\
 83.200$\pm$0.026  &   13.4  &  $l=0$, $n=10$ \\
 84.895$\pm$0.039  &    2.9  &  $l=3$?        \\
 86.275$\pm$0.030  &    6.3  &  $l=1$         \\
 87.284$\pm$0.025  &   18.2  &  $l=1$         \\
 88.407$\pm$0.028  &    8.6  &  $l=1$         \\
 89.951$\pm$0.026  &   13.7  &  $l=2$, $n=10$ \\
 90.864$\pm$0.025  &   17.9  &  $l=0$, $n=11$ \\
 91.677$\pm$0.035  &    3.9  &  $l=3$?        \\
 92.945$\pm$0.058  &    1.4  &  $l=3$?        \\
 93.400$\pm$0.031  &    5.6  &  $l=1$         \\
 94.697$\pm$0.025  &   17.4  &  $l=1$         \\
 95.785$\pm$0.025  &   15.1  &  $l=1$         \\
 97.636$\pm$0.026  &   12.5  &  $l=2$, $n=11$ \\
 98.427$\pm$0.024  &   19.7  &  $l=0$, $n=12$ \\
 99.510$\pm$0.033  &    4.7  &  $l=3$?        \\
101.442$\pm$0.025  &   16.2  &  $l=1$         \\
102.563$\pm$0.025  &   17.4  &  $l=1$         \\
104.284$\pm$0.027  &   10.4  &  $l=1$         \\
105.214$\pm$0.027  &   10.1  &  $l=2$, $n=12$ \\
105.851$\pm$0.026  &   12.6  &  $l=0$, $n=13$ \\
108.730$\pm$0.029  &    7.2  &  $l=1$         \\
110.116$\pm$0.025  &   14.0  &  $l=1$         \\
111.925$\pm$0.029  &    7.4  &  $l=1$         \\
112.846$\pm$0.027  &   10.2  &  $l=2$, $n=13$ \\
116.790$\pm$0.028  &    8.3  &  $l=1$         \\
118.052$\pm$0.027  &    9.5  &  $l=1$         \\
120.570$\pm$0.029  &    7.0  &  $l=2$, $n=14$ \\
121.284$\pm$0.029  &    6.8  &  $l=0$, $n=15$ \\
123.474$\pm$0.048  &    2.0  &  $l=1$         \\
125.560$\pm$0.031  &    5.8  &  $l=1$         \\
127.346$\pm$0.030  &    6.4  &  $l=1$         \\
132.901$\pm$0.034  &    4.3  &  $l=1$         \\
\hline
\noalign{\smallskip}
\end{tabular}
\label{tab.2}
\end{table}

\clearpage

\begin{table}
\caption{Results from BaSTI grid modeling, with uncertainties from the model calculations.
}
\begin{tabular}{cccc}
\hline
\noalign{\smallskip}
KIC     & M$_{\rm mo}$(M$_{\odot}$) & R$_{\rm mo}$(R$_{\odot}$) & T$_{\rm mo}$ (K) \\
\noalign{\smallskip} \hline
        &                         &                           &                    \\
\vspace{1mm}
9716522 & 2.33$_{-0.06}^{+0.11}$  & 12.21$_{-0.13}^{+0.16}$   & 4886$_{-39}^{+39}$ \\
\vspace{1mm}
9409513 & 2.35$_{-0.07}^{+0.07}$  & 10.61$_{-0.11}^{+0.13}$   & 4990$_{-39}^{+39}$ \\ 
\vspace{1mm}
9532903 & 2.21$_{-0.07}^{+0.07}$  &  9.00$_{-0.13}^{+0.10}$   & 5003$_{-39}^{+52}$ \\ 
\vspace{1mm}
9776739 & 2.21$_{-0.07}^{+0.06}$  &  8.91$_{-0.16}^{+0.10}$   & 4990$_{-39}^{+26}$ \\
\vspace{1mm}
9655101 & 2.23$_{-0.09}^{+0.07}$  &  8.75$_{-0.17}^{+0.09}$   & 4990$_{-39}^{+13}$ \\
\vspace{1mm}
9655167 & 2.15$_{-0.05}^{+0.05}$  &  8.49$_{-0.09}^{+0.10}$   & 5003$_{-26}^{+26}$ \\
\vspace{1mm}
9534041 & 2.11$_{-0.01}^{+0.01}$  &  8.24$_{-0.02}^{+0.06}$   & 4977$_{-00}^{+52}$ \\
9716090 & 2.10$_{-0.00}^{+0.00}$  &  8.07$_{-0.00}^{+0.00}$   & 4990$_{-00}^{+00}$ \\
        &                         &                           &                    \\
\hline
\noalign{\smallskip}
\end{tabular}
\label{tab.3}
\end{table}

\begin{table*}
\caption{Mode visibilities; the boldface value is $\left(V^{2}_{l=1}/V^{2}_{l=0}\right)_{\rm norm}$.}
\begin{tabular}{cccc}
\hline
\noalign{\smallskip}
KIC     & $\frac{V^{2}_{l=1}}{V^{2}_{l=0}}$ & $\frac{{V^{2}_{l=1}}/V^{2}_{l=0}}{\,\,\,\,\,\,\,\,\,\,\left(V^{2}_{l=1}/V^{2}_{l=0}\right)_{\rm norm}}$  &  $\frac{V^{2}_{l=2}}{V^{2}_{l=0}}$ \\
\noalign{\smallskip} \hline
        &                               &                           &                   \\
\vspace{1mm}
9716522 &    1.32$\pm$0.04              &  0.84$\pm$0.11            &   0.67$\pm$0.03   \\
\vspace{1mm}
9409513 &    1.36$\pm$0.03              &  0.86$\pm$0.11            &   0.64$\pm$0.04   \\
\vspace{1mm}
        &                            &                           &                    \\
9532903 &    1.62$\pm$0.24              &  1.02$\pm$0.20            &   0.67$\pm$0.06   \\
\vspace{1mm}
9776739 &    \underline{1.54$\pm$0.16}  &  0.98$\pm$0.16            &   0.59$\pm$0.06   \\
\vspace{1mm}
        &    {\bf 1.58$\pm\bf0.20$}        &                           &                   \\
\vspace{1mm}
        &                               &                           &                    \\
\vspace{1mm}
9655101 &    0.87$\pm$0.12              &  0.55$\pm$0.10            &   0.59$\pm$0.05    \\
\vspace{1mm}
9655167 &    0.86$\pm$0.09              &  0.55$\pm$0.09            &   0.63$\pm$0.04    \\
\vspace{1mm}
        &                               &                           &                    \\
9534041 &    1.14$\pm$0.05              &  0.72$\pm$0.10            &   0.60$\pm$0.01    \\
\vspace{1mm}
9716090 &    1.11$\pm$0.05              &  0.70$\pm$0.09            &   0.82$\pm$0.04    \\
\hline
\noalign{\smallskip}
\end{tabular}
\label{tab.4}
\end{table*}


\begin{thebibliography}{}
\bibitem[Angulo et al. (1999)]{1999NuPhA.656....3A} Angulo, C., Arnould, M., Rayet, M. et al.\ 1999, Nuclear Physics A, 656, 3 
\bibitem[Ballot et al. (2011)]{2011A&A...531A.124B} Ballot, J., Barban, C., and van't Veer-Menneret, C.\ 2011, A\&A, 531, A124
\bibitem[Bedding et al. (2011)]{2011Nature..471..608} Bedding, T.~R., Mosser, B., Huber, D. et al.\ 2011, \nat, 471, 608 
\bibitem[Borucki et al. (2010)]{2010Sci...327..977B} Borucki, W.~J., Koch, D., Basri, G. et al.\ 2010, Science, 327, 977
\bibitem[Brogaard et al. (2016)]{2016AN..XX..XX} Brogaard, K., Jessen-Hansen, J., Handberg, R. et al.\ 2016, AN special issue {\it Reconstruction of the Milky Way's History: Spectroscopic surveys, Asteroseismology and Chemo-dynamical models}, Guest Editors C. Chiappini, J. Montalb\'an, and M. Steffen, in press   
\bibitem[Buysschaert et al. (2016)]{2016A&A...588A..82B} Buysschaert, B., Beck, P.~G., Corsaro, E. et al.\ 2016, A\&A, 588, A82
\bibitem[Cantiello et al. (2016)]{2016arXiv160203056C} Cantiello, M., Fuller, J. and Bildsten, L.\ 2016, \apj, 824, 14 
\bibitem[Cassisi (2005)]{2005astro.ph..6161C} Cassisi, S.\ 2005, eprint arXiv:astro-ph/0506161
\bibitem[Chaplin \& Miglio (2013)]{2013ARA&A..51..353C} Chaplin, W.~J., Miglio, A.\ 2013, \araa, 51, 353
\bibitem[Christensen-Dalsgaard et al. (2008)]{2008JPhCS.118a2039C} Christensen-Dalsgaard, J., Arentoft, T., Brown, T.~M. et al.\ 2008, J. Phys. Conf. Ser. 118, 012039
\bibitem[Christensen-Dalsgaard (2011)]{jcd11} Christensen-Dalsgaard, J.\ 2011, arXiv:1106.5946v1
\bibitem[Constantino et al. (2015)]{2015MNRAS..452..123} Constantino, T., Campbell, T.~W., Christensen-Dalsgaard, J. et al.\ 2015, \mnras, 452, 123
\bibitem[Corsaro et al. (2012)]{2012ApJ..757..190} Corsaro, E., Stello, D., Huber, D. et al.\ 2012, \apj, 757, 190
\bibitem[Freytag et al. (1996)]{freytag96} Freitag, B., Ludwig, H.~G., \& Steffen, M.\ 1996, A\&A, 313, 497
\bibitem[Fergusen et al. (2005)]{2005ApJ..623..585F} Ferguson, J.~W., Alexander, D.~R., Allard, F. et al.\ 2005, \apj, 623, 585
\bibitem[Formicola et al. (2004)]{2004PhLB..591...61F} Formicola, A., Imbriani, G., Costantini, H. et al.\ 2004, Physics Letters B, 591, 61
\bibitem[Fuller et al. (2015)]{2015Sci...350..423F} Fuller, J., Cantiello, M., Stello, D. et al.\ 2015, Science, 350, 423
\bibitem[Girardi (1999)]{1999MNRAS.308..818G} Girardi, L.\ 1999, \mnras, 308, 818
\bibitem[Gough (1986)]{gough1986} Gough D. O.\ 1986, in Osaki, Y., ed., Hydrodynamic and magnetodynamic problems in the Sun and stars. University Tokyo Press, Tokyo, p. 117 
\bibitem[Grevesse \& Sauval (1998)]{1998SSRV..85..161G} Grevesse, N., Sauval, A.~J.\ 1998, Space Science Reviews, 85, 161
\bibitem[Handberg et al. (2017)]{handberg16} Handberg, R., Brogaard, K., Miglio, A., et al.\ 2017, \mnras, submitted
\bibitem[Handberg \& Lund (2014)]{2014MNRAS.445.2698H} Handberg, R., Lund, M.~N.\ 2014, \mnras, 445, 2698
\bibitem[Hekker et al. (2011)]{2011A&A...530A.100H} Hekker, S., Basu, S., Stello, D. et al.\ 2011, A\&A, 530, A100 
\bibitem[Hj{\o}rringgaard et al. (2017)]{hjg17} Hj{\o}rringgaard, J.~G., Silva Aguirre, V., White, T.~R. et al.\ 2017, \mnras, 464, 3713
\bibitem[Huber et al. (2010)]{2010ApJ...723.1607H} Huber, D., Beddring, T.~R., Stello, D. et al.\ 2010, \apj, 723, 1607 
\bibitem[Huber et al. (2011)]{2011ApJ..743..143} Huber, D., Bedding, T.\,R., Stello, D. et al.\ 2011, \apj, 743, 143
\bibitem[Iglesias \& Rogers (1996)]{1996ApJ..464..943I} Iglesias, C.~A., Rogers, F.~J.\ 1996, \apj, 464, 943
\bibitem[Kallinger et al. (2014)]{2014A&A...570A..41K} Kallinger, T., De Ridder, J., Hekker, S., et al.\ 2014, A\&A, 570, A41
\bibitem[Kippenhahn, Weigert \& Weiss (2012)]{2012sse..book.....K} Kippenhahn, R., Weigert, A., Weiss, A.\ 2012, Stellar Structure and Evolution (Springer-Verlag Berlin Heidelberg, 2012)
\bibitem[Koch et al. (2010)]{2010ApJ...713L..79K} Koch, D.~G., Borucki, W.~J., Basri, G., et al.\ 2010, \apjl, 713, L79
\bibitem[Magic et al. (2010)]{2010ApJ...718.1378M} Magic, Z., Serenelli, A., Weiss, A. and Chaboyer, B.\ 2010, \apj, 718, 1378
\bibitem[Miglio et al. (2012)]{2012MNRAS..419..2077} Miglio, A., Brogaard, K., Stello, D. et al.\ 2012, \mnras, 419, 2077
\bibitem[Molenda-\.{Z}akowicz et al. (2013)]{2013MNRAS.434.1422M} Molenda-\.{Z}akowicz, J., Sousa, S.~G., Frasca, A. et al.\ 2013, \mnras, 434, 1422 
\bibitem[Molenda-\.{Z}akowicz et al. (2014)]{2014MNRAS.445.2446M} Molenda-\.{Z}akowicz, J., Brogaard, K., Niemczura, E. et al.\ 2014, \mnras, 445, 2446 
\bibitem[Montalb{\'a}n et al. (2013)]{2013ApJ..766..118}  Montalb{\'a}n, J., Miglio, A., Noels, A. et al.\ 2013, \apj, 766, 118
\bibitem[Mosumgaard (2014)]{mosumgaard14} Mosumgaard, J.~R.\ 2014, Bachelor project, University of Aarhus
\bibitem[Mosser et al. (2010)]{2010A&A...517A..22M} Mosser, B., Belkacem, K., Goupil, M.-J. et al.\ 2010, A\&A, 517, A22  
\bibitem[Mosser et al. (2011)]{2011AA...532..A86} Mosser, B., Barban, C., Montalb{\'a}n, J. et al.\ 2011, A\&A, 532, A86   
\bibitem[Mosser et al. (2012a)]{2012A&A...537A..30M} Mosser, B., Elsworth, Y., Hekker, S. et al.\ 2012a, A\&A, 537, A30
\bibitem[Mosser et al. (2012b)]{2012A&A...540A..143M} Mosser, B., Goupil, M.~J., Belkacem, K. et al.\ 2012b, A\&A, 540, A143
\bibitem[Mosser et al. (2016)]{NewMosser} Mosser, B., Belkacem, K., Pin\c con, C. et al.\ 2016, in prep.  
\bibitem[Pietrinferni et al. (2004)]{basti04} Pietrinferni, A., Cassisi, S., Salaris, M. et al.\ 2004, \apj, 612, 168
\bibitem[Pinsonneault et al. (2014)]{2014ApJS..215...19P} Pinsonneault, M.\,H., Elsworth, Y., Epstein, C. et al.\ 2014, \apjs, 215, 23
\bibitem[Rogers \& Nayfonov (2002)]{2002ApJ..576..1064R} Rogers, F.~J., Nayfonov, A.\ 2002, \apj, 576, 1064 
\bibitem[Rogers, Swenson \& Iglesias (1996)]{1996ApJ..456..902R} Rogers, F.~J., Swenson, F.~J., Iglesias, C.~A.\ 1996, \apj, 456, 902  
\bibitem[Sandquist et al. (2016)]{Sand2016..831..11S} Sandquist, E., Jessen-Hansen, J., Shetrone, M.~D. et al.\ 2016, \aj, 831, 36 
\bibitem[Silva Aguirre et al. (2011)]{vsa11} Silva Aguirre, V., Ballot, J., Serenelli, A.~M, \& Weiss, A.\ 2011, A\&A, 529, 63
\bibitem[Silva Aguirre et al. (2015)]{vsa15} Silva Aguirre, V., Davies, G.~R., Basu, S. et al.\ 2015, \mnras, 452, 2127
\bibitem[Silva Aguirre et al. (2017)]{vsa17} Silva Aguirre, V., Lund, M.~N., Anita, H.~M. et al.\ 2017, \apj, accepted
\bibitem[Stello et al. (2009)]{2009MNRAS.400L..80S} Stello, D., Chaplin, W.~J., Basu, S. et al.\ 2009, \mnras, 400, L80
\bibitem[Stello et al. (2011)]{2011ApJ...739...13S} Stello, D., Meibom, S., Gilliland, R.~L.\ 2011, \apj, 739, 13 
\bibitem[Stello (2012)]{2012ASPC..462..200S} Stello, D.\ 2012, in Astronmical Society of the Pacific Conference Series, Vol. 462, Progress in Solar/Stellar Physics with Helio- and Asteroseismology, ed. H. Shibahashi, M. Takata \& A.~E. Lynas-Gray, 200
\bibitem[Stello et al. (2013)]{2013ApJ...765L..41S} Stello, D., Huber, D., Bedding, T.~R. et al.\ 2013, \apjl, 765, L41
\bibitem[Stello et al. (2016a)]{2016Natur.529..364S} Stello, D., Cantiello, M., Fuller, J. et al.\ 2016, Nature, 529, 364
\bibitem[Stello et al. (2016b)]{2016PASA...33...11S} Stello, D., Cantiello, M., Fuller, J. et al.\ 2016, \pasa, 33, 6 
\bibitem[Tassoul (1980)]{tassoul1980} Tassoul M.\ 1980, ApJS 43, 469
\bibitem[Vandakurov (1967)]{vandakurov1967} Vandakurov Y.~V.\ 1967, AZh. 44, 786
\bibitem[Weiss \& Schlattl (2008)]{2008ApSS..316..99W} Weiss, A., Schlattl, H.\ 2008, \apss, 316, 99
\end{thebibliography}
\end{document}